\documentclass[preprint,floats,floatfix,12pt,numreferences]{kluwer}
\usepackage{times} 
\usepackage[usenames,dvips]{color}
\usepackage{graphicx}   
\DeclareGraphicsExtensions{.pdf}
\graphicspath{{.},{./Fig_pdf/}} 
\newcommand{\mylab}[1]{\label{#1}}
\usepackage{ulem}

\newcounter{countuwe}

\newcounter{countcoaut}

\begin{document}
\begin{opening}
\title{The relation of steady evaporating drops fed by an influx and freely evaporating drops}
\author{Desislava \surname{Todorova} \email{d.todorova@lboro.ac.uk}}
%
\institute{Department of Mathematical Sciences, Loughborough University,
Loughborough, Leicestershire, LE11 3TU, UK}
\author{Uwe \surname{Thiele}\email{u.thiele@lboro.ac.uk}}
%
\institute{Department of Mathematical Sciences, Loughborough University,
Loughborough, Leicestershire, LE11 3TU, UK}
\author{Len M. \surname{Pismen}\email{pismen@technion.ac.il}}

\institute{Department of Chemical Engineering and Minerva Center for
Nonlinear Physics of Complex Systems, Technion--Israel Institute of
Technology, 32000 Haifa, Israel}
\begin{abstract}
  We discuss a thin film evolution equation for a wetting evaporating
  liquid on a smooth solid substrate. The model is valid for slowly
  evaporating small sessile droplets when thermal effects are
  insignificant, while wettability and capillarity play a major role.
  The model is first employed to study steady evaporating drops that are fed
  locally through the substrate. An
  asymptotic analysis focuses on the precursor film and the transition
  region towards the bulk drop and a numerical continuation of steady
  drops determines their fully non-linear profiles.

  Following this, we study the time evolution of freely evaporating
  drops without influx for several initial drop shapes.  As a result
  we find that drops initially spread if their initial contact angle
  is larger than the apparent contact angle of large steady
  evaporating drops with influx. Otherwise they recede right from the
  beginning.
\end{abstract}
\end{opening}
%
\section{Introduction}
\mylab{sec:intro}
%
Evaporation of thin liquid films and sessile droplets has attracted
much attention both as the way to probe the dynamics of the
contact line \cite{SLT98,LLL03} and as a route to create deposition
patterns through sedimentation of solutes and suspensions
\cite{Deeg97,TMP98,Deeg00,Huan05}. A number of studies concentrate on
problems pertinent to any evaporation process, which include mass and
heat transfer and thermocapillarity \cite{AnDa95,Ajae05b}. For slowly
evaporating small sessile droplets studied in contemporary
well-controlled experiments on smooth surfaces
\cite{SLT98,CBC02}, thermal effects are, however, insignificant,
while contact line dynamics and liquid-substrate interactions play a
major role. It has been suggested that the relation between spreading and
evaporation/condensation goes both ways, so that the latter may
alleviate the notorious contact line singularity \cite{W93,P02}.
For background on spreading see, e.g., \cite{Bonn09,SVR07}.

A remarkable phenomenon observed in evaporating completely wetting
liquids is the formation of a dynamic meniscus with a finite contact
angle \cite{Deeg97,CBC02,noushineevap}. The standard approach to
computing the form of a spreading and evaporating drop and the
resulting dynamic contact angle \cite{AnDa95,Hock95,arezki} is based
on the lubrication approximation with the singularity at the contact
line alleviated by slip. The evaporation rate is treated in three
alternative ways. One possibility, realized in the presence of a
temperature difference between the substrate and vapor phase
\cite{AnDa95}, is the evaporation rate determined by the balance of
latent heat and heat flux. The evaporation rate is then uniform in the
limit of small Biot numbers, but diverges near the contact line in the
opposite limit. Both limits can also be realized under isothermal
conditions. The evaporation rate is uniform (as long as the layer
thickness remains outside the range of intermolecular forces) when
evaporation is controlled either by phase transition kinetics at the
interface or by diffusion through a boundary layer of constant
thickness in a stirred vessel. If evaporation is diffusionally
controlled with no stirring, the evaporation rate increases towards
the contact line; an analytical solution based on interfacial
equilibrium with no flux onto the unwetted substrate yields the flux
diverging on the contact line \cite{Deeg00b,arezki}. Another approach
accounts for the influence of the thermal conductivity of the
substrate and the dependence of the vapor saturation concentration on
temperature \cite{Dunn09,Sefi09}.

The aim of the present paper is to present a simple isothermal thin
film evolution equation with evaporation limited by phase transition
kinetics (or boundary layer transfer, but not diffusion), that
correctly describes the influence of effective molecular interactions
on evaporation in the case of complete wetting. This is achieved by
taking into account the dependence of the saturated vapor pressure on
the disjoining pressure and curvature in the way it has been done in
studies of the dynamics of evaporating films \cite{PKS99,LGP02,Pism04}
but not in the cited studies of droplet spreading. This allows us to
describe in a consistent way the transition from the bulk droplet to a
precursor layer and eliminate singularities at the contact line. Note
however, that our model may be obtained as the isothermal limit of the
models in \cite{Ajae05b,ReCo10}, i.e., letting the difference of substrate and
ambient temperature, and the latent heat go to zero. It also corresponds
to the limit of infinite thermal conductivity of the liquid.

Following \cite{AnDa95,Hock95}, we consider a two-dimensional ``fed''
system that allows us to study steady states of evaporating droplets.
These steady states are compared to droplet shapes resulting from a
time evolution of an evaporating droplet (without influx). The
comparison will be employed to discuss a possible special role the
steady state profiles play in the time evolution. A related approach
is taken in Ref.~\cite{AGS10} where steady fronts of evaporating
liquid on an incline are considered.

The paper is structured as follows. The following
section~\ref{sec:mod} introduces our model and discusses the scaling,
whereas section~\ref{sec:asymp} discusses the asymptotics in the
precursor film. Section~\ref{sec:ssd} discusses the properties of
steady drops with influx as a function of the influx
strength and of the single remaining dimensionless parameter. The
time-evolution in the case without influx is analysed in
section~\ref{sec:timesim} where we also compare the steady drop
profiles in the case with influx to the time-dependent profiles in the
non-fed case.  Section~\ref{sec:disconc} gives our conclusions.

\section{Basic Model and Scaling}
\mylab{sec:mod}

For simplicity, we restrict our attention to a two-dimensional system
as sketched in Fig.~\ref{fig:sketch}. Conceptually, there exists no
difference to the full three-dimensional system, we only expect the
transport rates to change.
\begin{figure}[tbh]
\includegraphics[width=1.0\hsize]{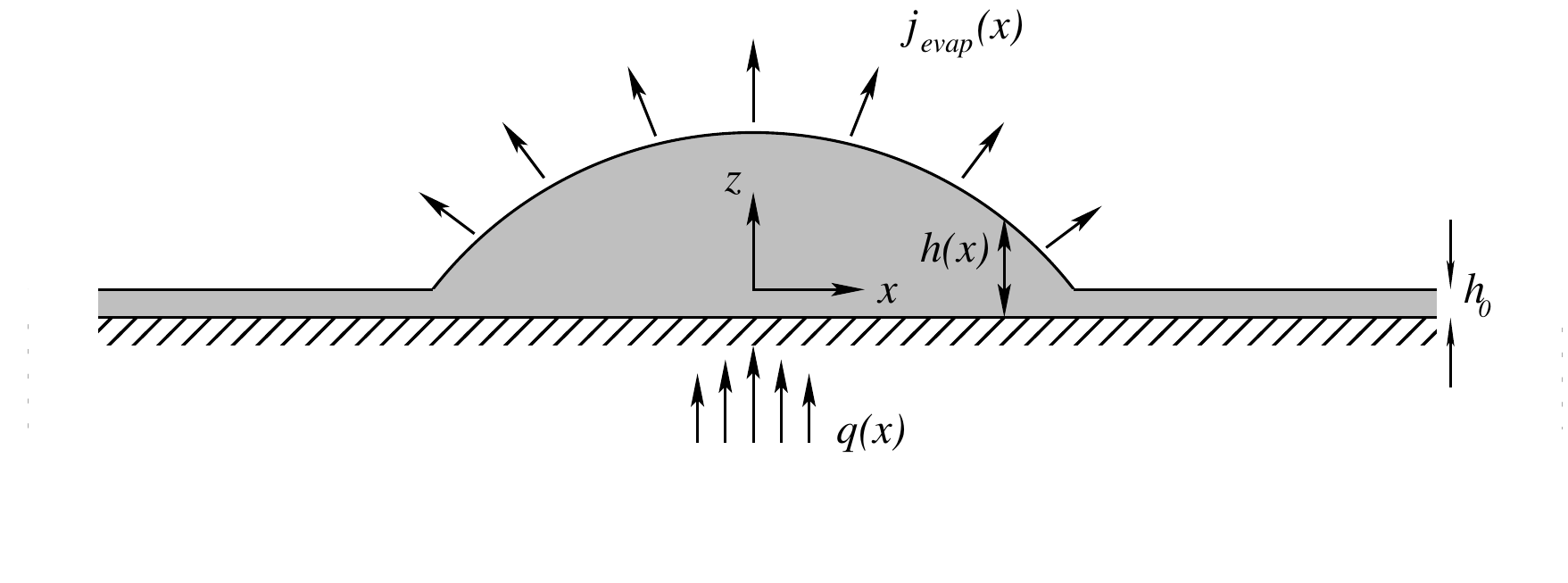}
\caption{Sketch of the two-dimensional geometry employed for investigating  an
  evaporating droplet with localised influx $q(x)$.}
\mylab{fig:sketch}
\end{figure}

Using the lubrication approximation, the evaporation dynamics for an
isothermal droplet of liquid on a porous substrate is captured by an
evolution equation for the film thickness profile $h$
\cite{ODB97,LGP02,Pism04,Thie10}
\begin{eqnarray}
\partial_t\,h &=& - \partial_x\,j_\mathrm{conv}(x) - j_\mathrm{evap}(x) 
+ q(x) , 
\mylab{film} \\
\nonumber
j_\mathrm{conv}(x) &=& - \frac{h^3}{3\eta}\,\partial_x p, 
\qquad
j_\mathrm{evap}(x)=\beta\,\left(\frac{p}{\rho} - \mu_0\right),\\
p &=& - \gamma \, \partial_{xx} h - \Pi(h) .
\mylab{flux}
\end{eqnarray}
The first and second term on the r.h.s.~of Eq.~(\ref{film}) are the
divergence of the convective flux $j_\mathrm{conv}(x)$ and the
evaporative flux $j_\mathrm{evap}(x)$ , which correspond,
respectively, to the conserved and non-conserved part of the dynamics.
The function $q(x)$ is the influx through the (locally) porous
substrate.  The evaporative flux is proportional to the difference
between the chemical potential of the ambient vapour and the chemical
potential in the liquid $\mu=p/\rho$; $p$ is pressure, $\beta$ is an
effective evaporation rate constant; and $\gamma$, $\rho$ and $\eta$
are the surface tension, mass density and dynamic viscosity of the
liquid, respectively. The pressure $p$ contains the curvature pressure
$-\gamma\partial_{xx} h$ and the disjoining pressure $\Pi(h)$
modelling wettability \cite{deGe85,Isra92,ODB97,KaTh07,Bonn09}; the
hydrostatic pressure is neglected as we focus on nano- and
micro-droplets.

To model a droplet of completely wetting liquid, we employ a
long-range stabilising van der Waals disjoining pressure $\Pi=-A/6\pi
h^3$ with the Hamaker constant $A<0$
\cite{RuJa74,Isra92,Shar93,LGP02,Thie07,Thie10}. Note that other sign
conventions are also common (cf.~e.g.~\cite{deGe85,StVe09,Bonn09}).
The model~(\ref{film}) is related to various models in the
literature: it may be obtained from the one in \cite{LGP02} by adding
an influx and replacing the disjoining pressure for a partially
wetting liquid by one for a wetting liquid. The models in
\cite{Ajae05b,ReCo10} incorporate various thermal aspects that are
here neglected by assuming that the latent heat is very small or/and
the thermal conductivity is very large.
The same applies to the steady
state description in \cite{Morr01}. Note that our model also
corresponds to the one in \cite{ReCo10} in the limit of zero superheat.

A dimensionless form of Eqs.~(\ref{film}) and (\ref{flux}) can be
obtained by choosing the characteristic energy density of molecular
interactions between the fluid and substrate $\kappa=|A|/6\pi h_0^3$ as
the pressure scale and the equilibrium film thickness $h_0=|A/6\pi
\rho \mu_0|^{1/3}$ corresponding to the ambient vapour potential $\mu_0$
as the scale of film thickness $h$ (note that $\mu_0<0$ when a thick
flat film evaporates). The horizontal coordinate $x$ and time $t$ can be
scaled in several ways \cite{Pism04}. A short horizontal length scale
\begin{equation}
l=\sqrt{\frac{\gamma h_0}{\kappa}} 
= h_0^2\,\sqrt{\frac{6\pi\gamma}{|A|}} 
 = \left(\frac{|A|}{6\pi}\right)^{1/6}\frac{\sqrt{\gamma}}{|\rho \mu_0|^{2/3}}
\mylab{short}
\end{equation}
is fixed by the balance between disjoining pressure and surface
tension at the thickness of the wetting layer, and determines the
extent of a region adjacent to the contact line where the interface
may be strongly curved due to interaction with the
substrate. The lubrication approximation remains formally
  applicable as long as $l$ far exceeds $h_0$. Note, however, that
  lubrication approximation often still predicts the qualitative
  behavior for many systems with larger contact angles
  \cite{ODB97,KaTh07}. When considering the results obtained with
  models like Eqs.~(\ref{film}), one has always to keep in mind that
  even very large contact angles obtained in lubrication
  approximation (measured as slopes at the inflection point of the
  drop profiles) correspond to rather small angles in physical scaling.

Another horizontal scale, applicable in the precursor layer, is
determined by the balance of flow driven by the disjoining pressure
gradient and evaporation:
\begin{equation}
L=\sqrt{\frac{h_0^3 \rho}{\beta \eta}}
=\sqrt{\left|\frac{A}{6\pi \mu_0\beta \eta}\right|}, 
\mylab{long}
\end{equation}
This scale is large when evaporation is slow. 
It is appropriate to choose $L$ as the horizontal scale, assuming
it to be of the same order of magnitude as the third available scale
-- the droplet size. The respective time scale is $T=(L/h_0)^2 \eta/
\kappa$, and the flux $j_\mathrm{conv}$ is scaled by $h_0L/T$. 

Retaining the same notation for the rescaled variables, we
rewrite Eqs.~(\ref{film}) and (\ref{flux}) as
\begin{eqnarray}
\partial_t\,h &=& - \partial_x\,j_\mathrm{conv}(x) + \left(\epsilon \,\partial_{xx} h\,+\,\frac{1}{h^3} - 1 \right) + q(x) , 
\mylab{film:dl} \\
j_\mathrm{conv}(x) &=& \frac{h^3}{3}\partial_x\,\left(\epsilon\, \partial_{xx} h\,+\,\frac{1}{h^3}\right), 
\mylab{flux:dl} \end{eqnarray}
where the parameter 
\begin{equation}
\epsilon=\frac{(6\pi)^{2/3}\gamma\beta\eta}{|A|^{2/3}\rho^{4/3}|\mu_0|^{1/3}}
\mylab{eq:eps}
\end{equation}
denotes the scale ratio $(l/L)^2$. Note that it is proportional to the
evaporation rate constant $\beta$, but contains as well a weak
dependence on the chemical potential $\mu_0$ in the denominator.

In the following we will study steady state droplets that are obtained
for an influx $q(x)$ localised at the centre of the drop
(section~\ref{sec:ssd}). Below, the steady profiles are compared to
time simulations without influx for different initial profiles
(section~\ref{sec:timesim}).  First, however, we discuss the
asymptotics in the precursor film.
%
\section{Asymptotics in the precursor film}
\mylab{sec:asymp}
In the outer precursor region, the film is almost flat and surface
tension can be neglected, i.e., we set $\epsilon=0$ in
Eqs.~(\ref{film:dl}) and (\ref{flux:dl}) and assume $q(x)=0$. In the
linear regime the film thickness decays exponentially to its
equilibrium value $h=1$:
\begin{equation}
h-1\sim\exp(-\sqrt{3}x).
\mylab{eq:out-exp}
\end{equation}
Note the difference from the non-physical asymptotics $h \sim x^{1/4}$
in \cite{arezki} where  the dependence of the evaporation equilibrium 
on the disjoining pressure was neglected.  The latter profile 
corresponds to the well-known result of de Gennes \cite{deGe85} who
failed to recognise it as an \emph{unstable} solution.

To discuss the nonlinear behaviour we reduce Eqs.~(\ref{film:dl}) and
(\ref{flux:dl}) to the stationary equation
\begin{equation}
\frac{d^2\ln h}{dx^2} \,=\, 1 -\,\frac{1}{h^3} .
\mylab{film:out}
\end{equation}
This equation is solved by using $h$ as an independent variable, and
$y(h)=(d\ln h/dx)^2$ as a dependent variable. The transformed equation
is
\begin{equation}
y'( h) \,=\, \frac{2}{h} \left(1 -\,\frac{1}{h^3} \right).
\mylab{film:out1}
\end{equation}
It is integrated with the boundary condition $y(1)=0$ to yield 
\begin{equation}
y( h) \,=\,  2\left[ \ln h \, -\,\frac{1}{3}\left(1-\frac{1}{h^3}\right)\right].
\mylab{film:out2}
\end{equation}
The precursor film profile is obtained in an implicit form
\begin{equation}
\sqrt{2}x \,= 
\int \left[ H \, -\,\frac{1}{3}\left(1-e^{-3H}\right)\right]^{-1/2} dH,
\mylab{film:out3}
\end{equation}
where $H=\ln h$. 
This solution formally implies a very fast growth $h \sim \exp[(x-x_0)^2]$  
towards the bulk of the droplet. It becomes, however, inapplicable as $h$ grows, 
necessitating a modified scaling.
One can see that the two terms in the r.h.s.~of Eq.~(\ref{flux:dl}) become, up to logarithmic corrections,
comparable at $h \sim \epsilon^{-1/4}$, which, though appreciably exceeding
the thickness of the equilibrium wetting layer $h=1$, may be still far below the
height of the bulk droplet. 
As follows from Eq.~(\ref{film:out2}), the incline at this thickness
level is, up to logarithmic corrections,
$h_x =h \sqrt{y(h)} \sim \epsilon^{-1/4}$ in agreement with results by
Morris \cite{Morr01,Morr03}.
This sheds light on the origin of a finite contact angle in an
evaporating droplet. As we will see below in section~\ref{sec:ssd} the
numerically obtained dependence agrees well with the asymptotic
result.

The flux $J$ from the droplet bulk into the precursor at a
``transitional'' location $X$ corresponding to the thickness level 
$h=\epsilon^{-1/4}\zeta$ is
determined by the total evaporation rate from the precursor,
which can be obtained directly from Eq.~(\ref{film:out}):
\begin{equation}
J  = \int_X^\infty \left(1-\frac{1}{h(x)^3}\right) dx 
=  - \left(\frac{d\ln h}{dx} \right)_{x=X} 
\approx  \sqrt{2\left( \ln \frac{\zeta}{\epsilon^{1/4}} \, -\,\frac{1}{3}\right)}.
\mylab{film:js}
\end{equation}
The dependence both on $\epsilon$ and on a precise choice of the
level $\zeta$ is very weak. The rest of evaporation goes at an
almost constant rate from the bulk of a large droplet.

\section{Steady state droplets with influx}
\mylab{sec:ssd}

For zero influx through the porous substrate ($q(x)=0$) the only
steady state solution is $h=h_0$. However, for $q(x)\neq0$ steady
droplets may exist with a volume determined by the dynamic equilibrium
between the overall influx through the substrate and the overall
evaporation flux.

Here we use continuation techniques \cite{AUTO97,DKK91,DKK91b} to
numerically analyse the steady state solutions of Eqs.~(\ref{film:dl})
and (\ref{flux:dl}), i.e., we set $\partial_t\,h=0$ and solve the
resulting ordinary differential equation as a boundary value problem
on a domain of size $D$ with the boundary conditions (for a
symmetrical drop) $\partial_x h=\partial_{xxx} h=0$ at $x=0$ 
(drop centre). At $x=D$ we employ either $h=1$ and $\partial_{x} h=0$, or
$\partial_{x} h=0$ and $\partial_{xx} h=0$. If the domain is
sufficiently large the results depend neither on the particular
choice of $D$ nor on the used version of boundary conditions at $x=D$.
For details on the usage of continuation methods for thin film
equations see, e.g., Refs.~\cite{Thie02}, \cite{JBT05} and
\cite{ThKn06} where they have been employed to study sliding drops,
chemically driven running drops and drops pinned by wettability
defects, respectively.

For the influx $q(x)$ we use a normalised Gaussian 
\begin{equation}
q(x) = q_0\frac{2}{\sigma\sqrt{\pi}}\exp\left[ -\frac{x^2}{\sigma^2}\right]
\mylab{gauss}
\end{equation}
with $q_0=\int_0^\infty q(x)dx$ being the total influx through the
substrate. If the droplet size is large as compared to the width
$\sigma$, the results do not depend on the particular choice of $\sigma$.

\begin{figure}[htbp]
\includegraphics[width=0.49\hsize]{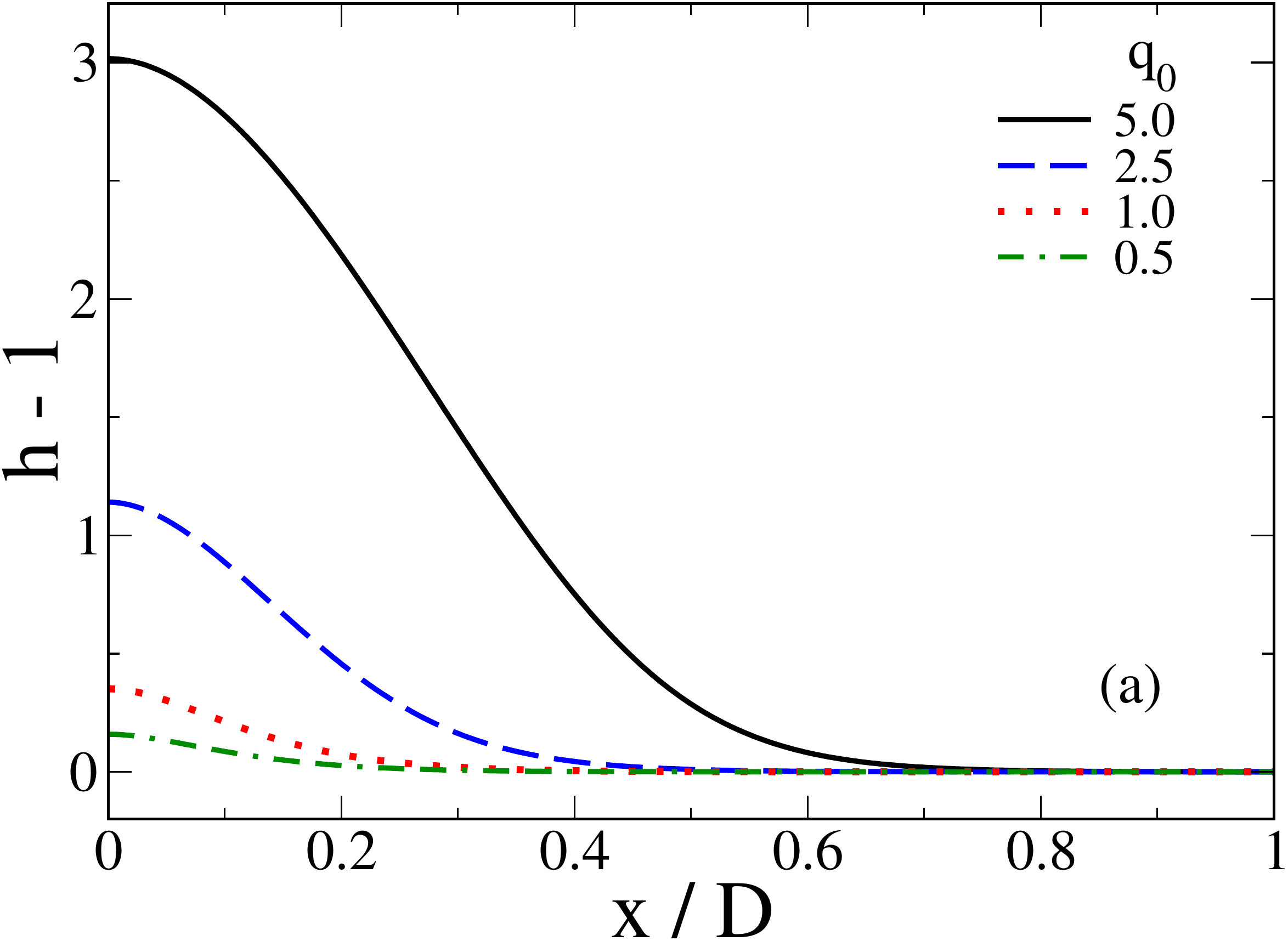}
\includegraphics[width=0.51\hsize]{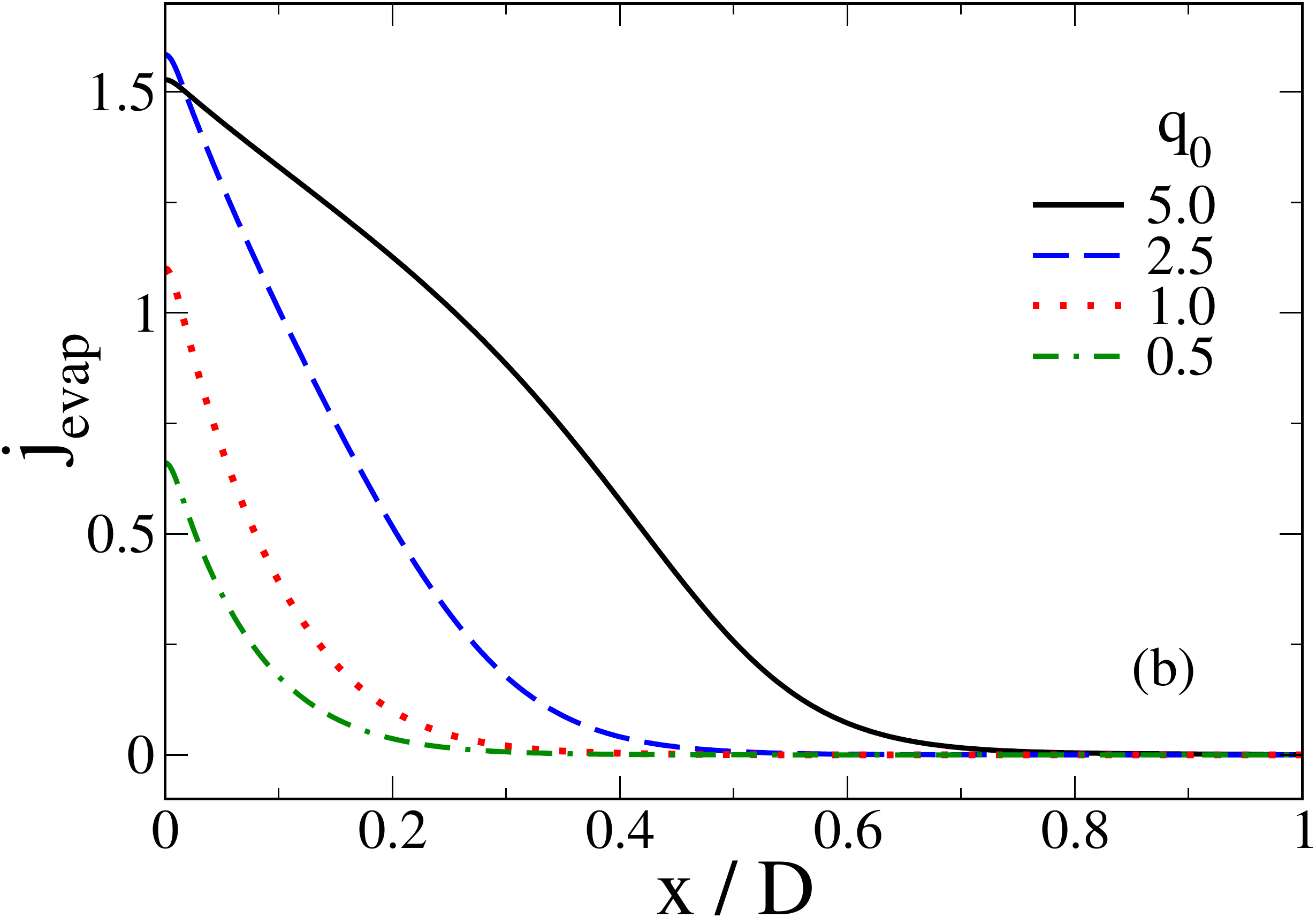}
\caption{(colour online) For the case of small droplets we give (a) droplet profiles
  and (b) evaporation flux in dependence of position for
  $\epsilon=1.0$ and various total influxes $q_0$ as given in the
  legend. Domain size is $D=10$, and $\sigma=0.1$.}
\mylab{fig:prof}
\end{figure}

\begin{figure}[htbp]
\includegraphics[width=0.49\hsize]{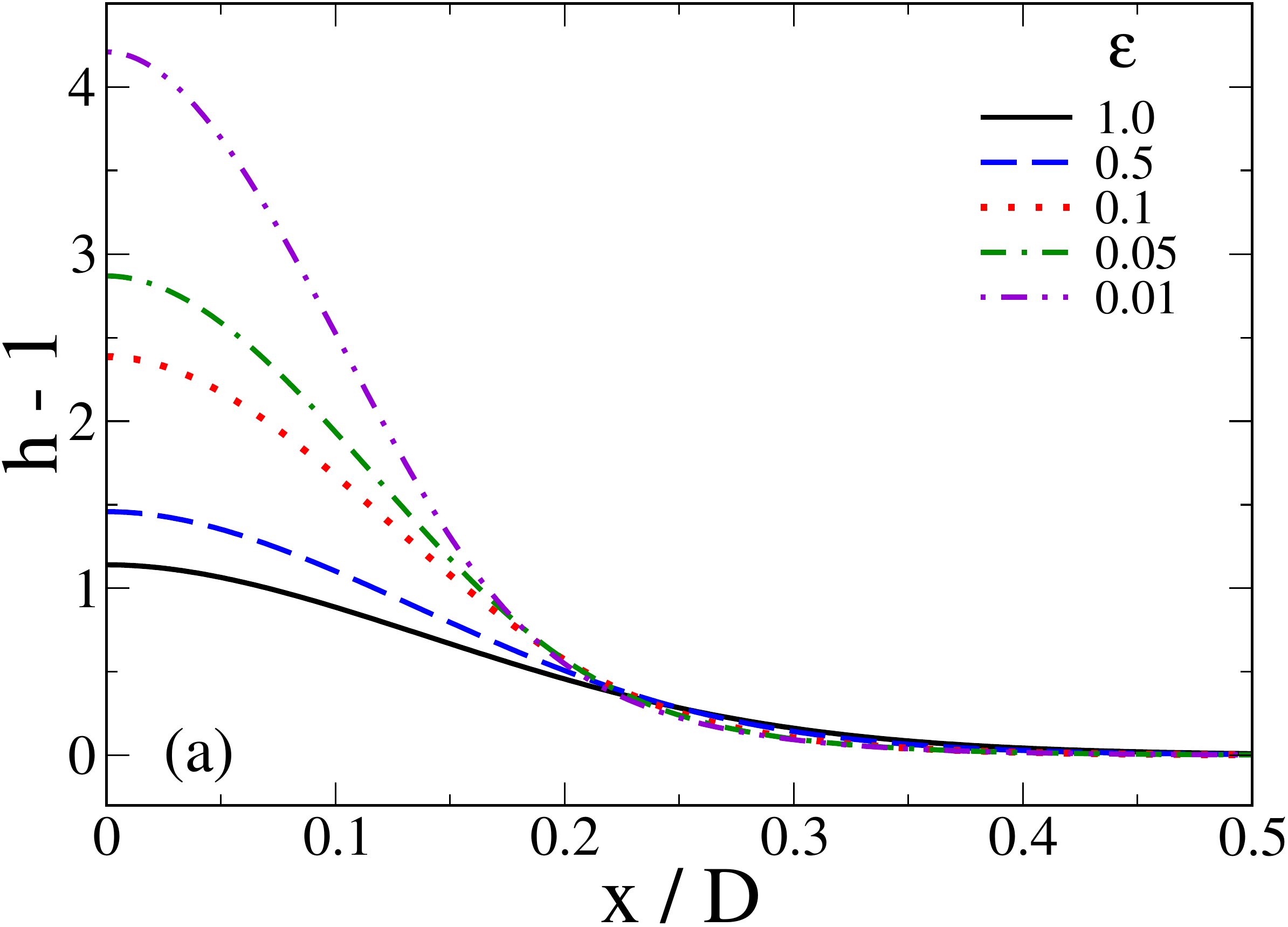}
\includegraphics[width=0.51\hsize]{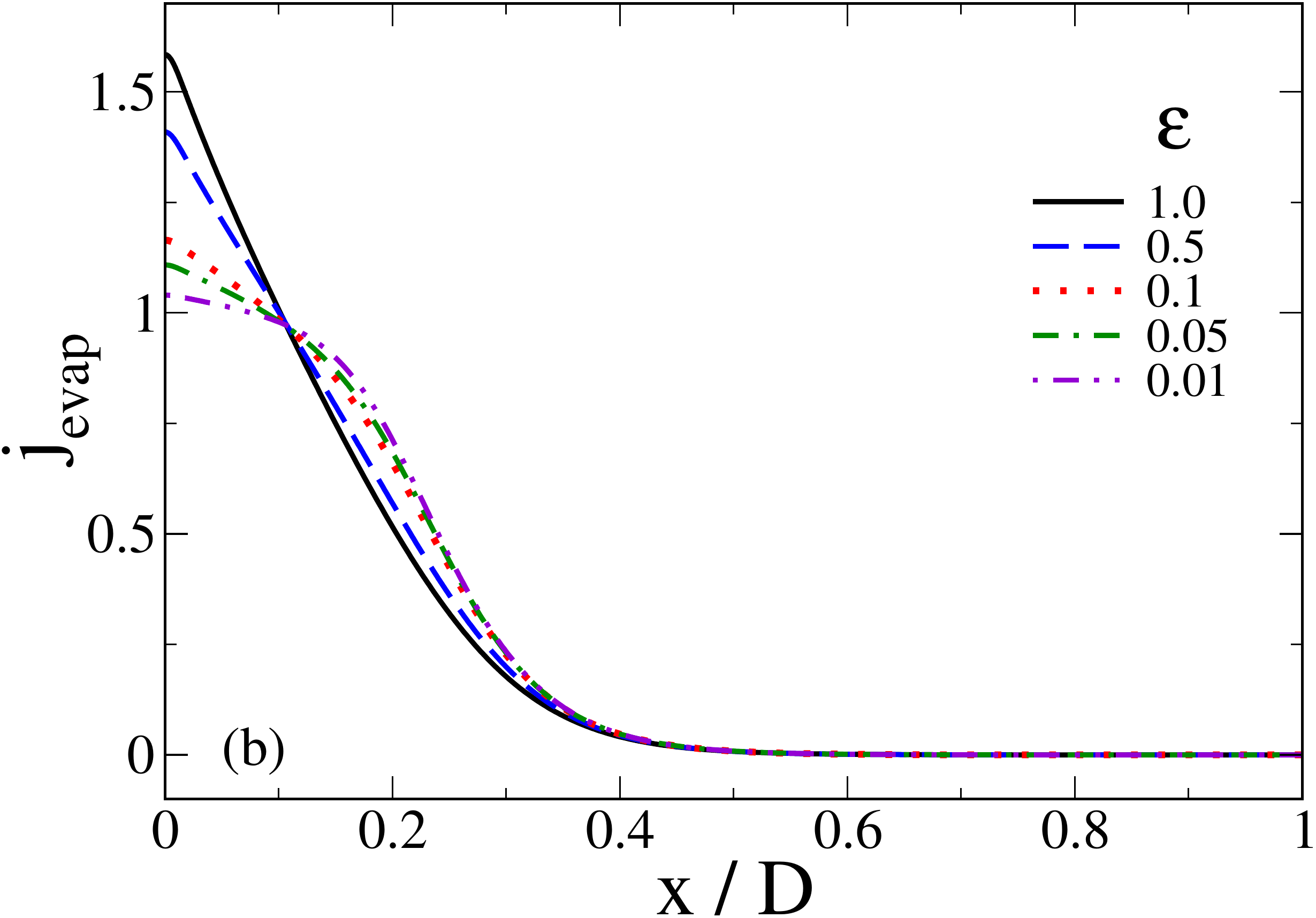}
\caption{(colour online) For the case of small droplets we show (a) droplet profiles
  and (b) evaporation flux in dependence of position for $q_0=2.5$ and
  various $\epsilon$ as given in the legend. Domain size is
  $D=10$, and $\sigma=0.1$.}
\mylab{fig:prof2}
\end{figure}

Figs.~\ref{fig:prof} and~\ref{fig:prof2} show profiles of
  rather small (nano-)droplets (left panels) and the corresponding
  dependencies of the evaporative flux on position (right panels). The
  results are given for various moderate values of the influx $q_0$
  (Fig.~\ref{fig:prof}) and the length scale ratio $\epsilon$
  (Fig.~\ref{fig:prof2}). In all shown cases these droplets are not
much higher than the wetting layer. For such small drops the behaviour
is dominated by the influence of the disjoining pressure. In
consequence, the evaporation decreases monotonically from the center
of the drops towards the contact region. Interestingly, in all cases
the droplet assumes a shape that does not allow for any condensation
of liquid even in the contact line region where the Laplace pressure
is negative. Note that there exists a one-to-one correspondence
  between the strength of the influx $q_0$ and droplet volume for
  fixed $\epsilon$. This implies that one may characterise the
  relative size of droplets either by volume or by influx $q_0$.

For extremely small drops (see, e.g., profile for $q_0=0.5$ in
Fig.~\ref{fig:prof}) the disjoining pressure influence is stronger
than the capillary pressure even at the drop centre. As a result, the
absolute value of the evaporation flux $j_\mathrm{evap}$is smaller
than one even at the centre of the drop. For slightly larger drops
(see, e.g., profile for $q_0=2.5$ in Fig.~\ref{fig:prof}) the
capillary pressure dominates the disjoining pressure at the drop
centre and $j_\mathrm{evap}$ is larger than one. With a further
increase in drop size the influence of the capillary pressure
diminishes and $j_\mathrm{evap}$ eventually approaches unity
everywhere with the exception of the contact line region
(cf.~Fig.~\ref{fig:prof3}).

\begin{figure}[htbp]
\includegraphics[width=0.5\hsize]{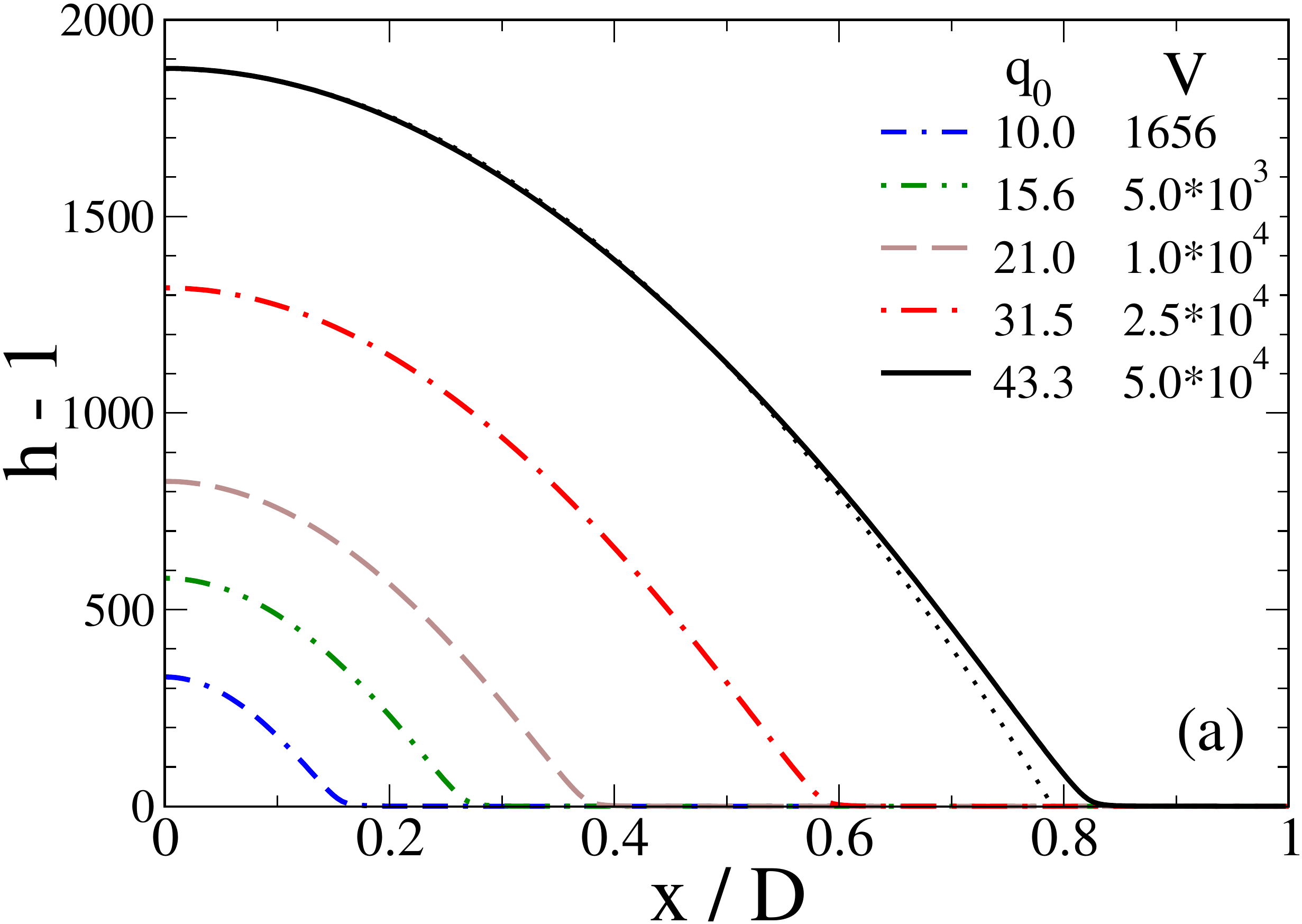}
\includegraphics[width=0.5\hsize]{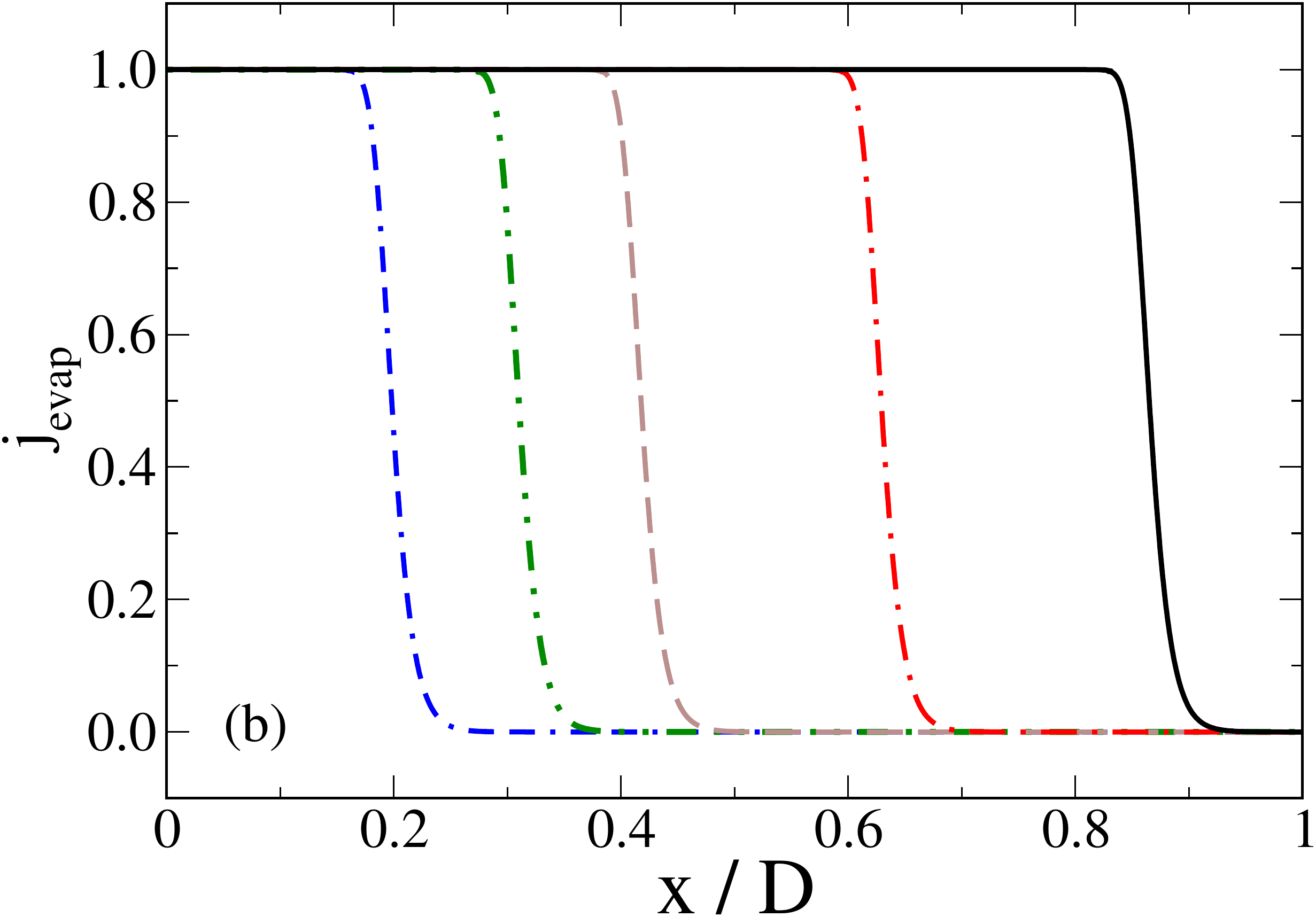}
\includegraphics[width=0.5\hsize]{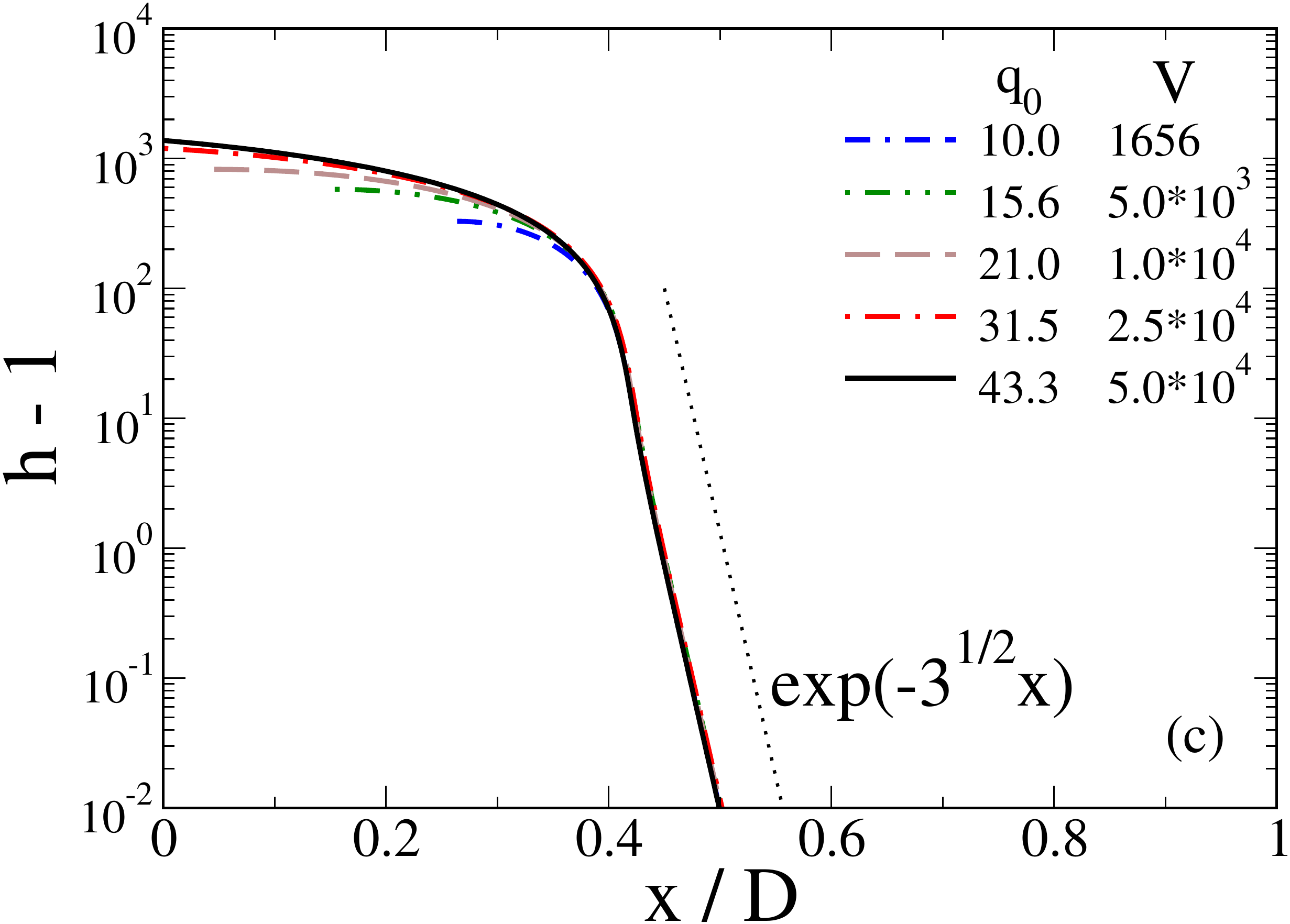}
\caption{(colour online) For the case of large drops we show for $\epsilon=10^{-6}$
  (a) drop profiles and (b) evaporation flux in dependence of
  position. Results are given for various total influxes $j_0$ and
  droplet volumes $V$ (see legend). The thin dotted line in (a) gives
  for the largest drop the corresponding parabolic drop profile of
  identical maximal height and curvature at centre (corresponding to a
  spherical cap in lubrication approximation).  Panel (c) shows $\log
  h$ to indicate the universal behaviour near the contact line (drops
  shifted in $x$). The dotted line indicates the linear result
  $h-1\sim\exp(-\sqrt{3}x)$ [Eq.~(\ref{eq:out-exp})].  Domain size is
  $D=50$, and $\sigma=0.1$.}
\mylab{fig:prof3}
\end{figure}

\begin{figure}[htbp]
\includegraphics[width=0.5\hsize]{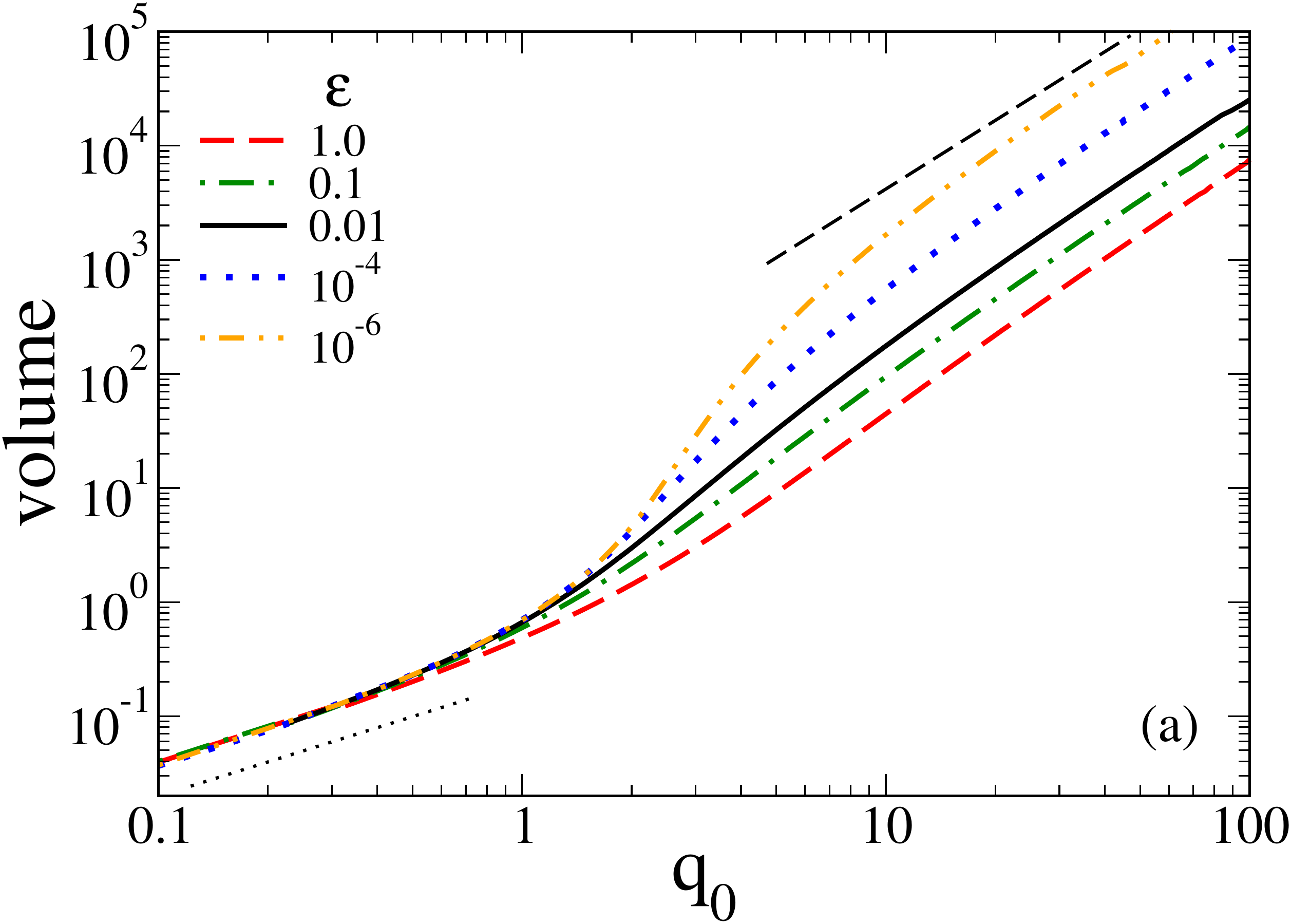}
\includegraphics[width=0.5\hsize]{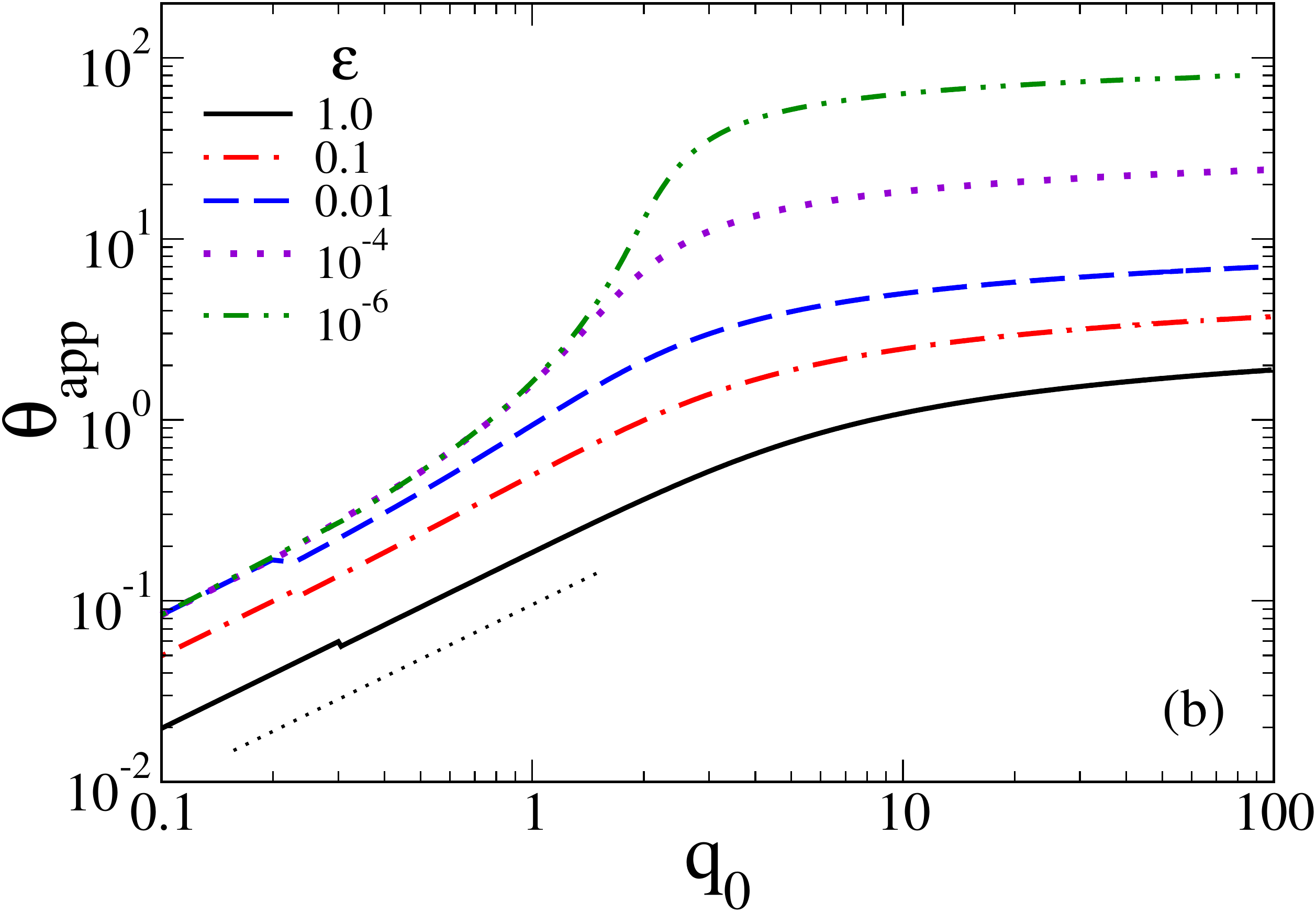}
\caption{Shown are (a) drop volume and (b) the
apparent contact angle $\theta_\mathrm{app}$ (defined as the maximal
slope of the drop profile), in dependence of total influx $q_0$ for
  various length scale ratios $\epsilon$ as given in the legend. The
  straight dotted [dashed] lines indicate linear [quadratic]
  dependencies, respectively.}
\mylab{fig:vol}
\end{figure}

Decreasing $\epsilon$ mainly influences the height of the droplets while
the width remains roughly constant [Fig.~\ref{fig:prof2}(a)].  This
implies that the curvature at the drop centre and the apparent contact
angle $\theta_\mathrm{app}$ (defined as the maximal
slope of the drop profile), increase with decreasing $\epsilon$. However, although curvature
increases we find that the influence of capillarity on the evaporation
flux decreases [Fig.~\ref{fig:prof2}(b)].  For $\epsilon=0.01$, one
has at the drop centre $j_\mathrm{evap}$ slightly above one.
Furthermore, at the same $\epsilon$, $j_\mathrm{evap}$ has already a
small plateau at the drop centre, i.e., the flux is nearly constant at
the value determined solely by the chemical potential.

The influence of the source width $\sigma$ is marginal as long as it
is sufficiently smaller than the droplet width.  For moderately large width
it has still no influence on the contact line region but has some
influence on the center of the drop.  Increasing, for instance,
$\sigma$ from 0.1 to 1.0 at constant $j_0=2.5$ and $\epsilon=1$ the
drop volume goes up by about 5\%. Decreasing $\sigma$ down to 0.001
has no visible influence on the droplet shape.

The droplets discussed up to this point represent nano-droplets of heights
normally below 500 nm. However, for much smaller $\epsilon$ or much
larger $q_0$ one is able to study micro-droplets with heights in the
10-100 $\mu$m range.  Fig.~\ref{fig:prof3} shows profiles of such drops
and the local evaporative flux for $\epsilon=10^{-6}$. For such large
drops the local evaporation is essentially constant for the 'bulk
drop' and decreases monotonically in a confined contact region
[Fig.~\ref{fig:prof3}(b)]. Panel (c) of Fig.~\ref{fig:prof3} shows the
logarithm of $h-1$. By shifting the drops in the $x$-direction one can
appreciate that the approach to $h=1$ is universal and well described
by the linear relation $h-1\sim\exp(-\sqrt{3}x)$ derived above (see
Eq.~(\ref{eq:out-exp}) in section~\ref{sec:asymp}).

\begin{figure}[htbp]
\includegraphics[width=0.5\hsize]{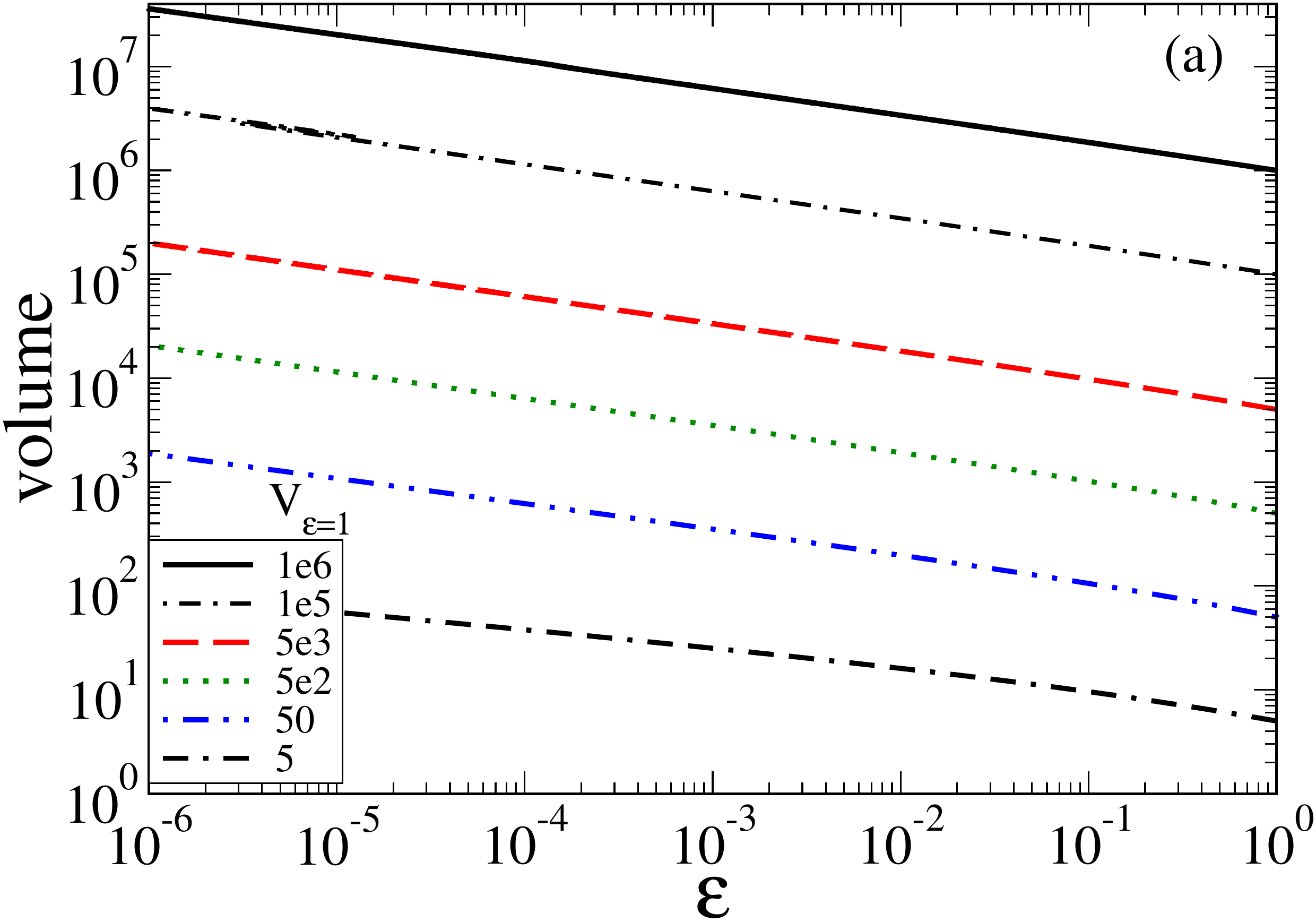}
\includegraphics[width=0.5\hsize]{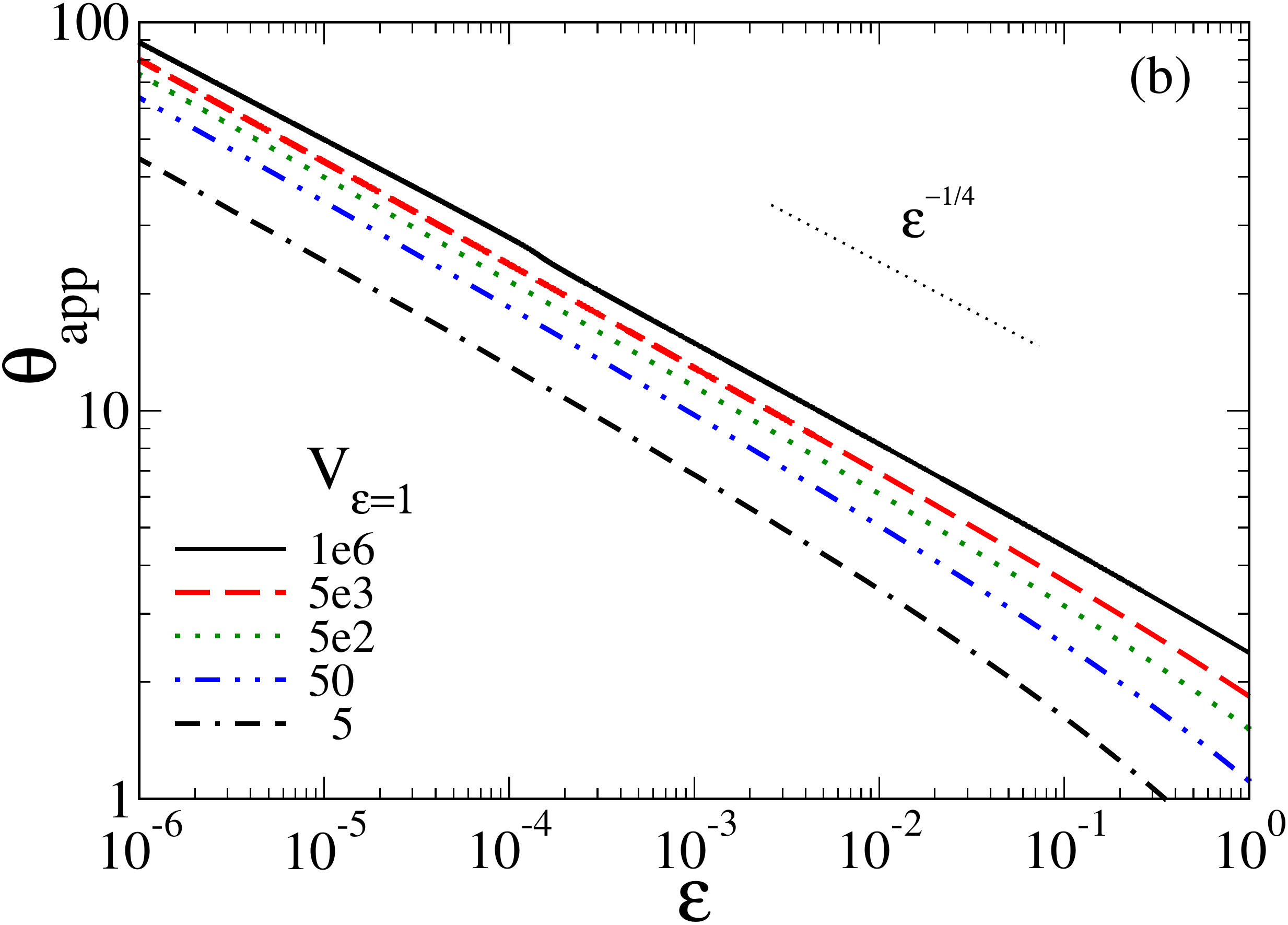}
\caption{(colour online) Shown are (a) drop volume and (b) apparent
  contact angle in dependence of the length scale ratio $\epsilon$.
  The total influx $q_0$ is constant for each line, respectively. The
  lines are characterised by the drop volume at $\epsilon=1.0$ (see
  legends).}
\mylab{fig:vol2}
\end{figure}

When increasing the influx for fixed $\epsilon$ the steady drops
become larger in width and height [Fig.~\ref{fig:prof}(a)]. This is
indicated as well by the dependence of volume on influx
[Fig.~\ref{fig:vol}(a)]. The corresponding apparent contact angle is
shown in Fig.~\ref{fig:vol}(b). One clearly distinguishes a small-drop
and large-drop regime with a crossover at about $V=1$. In the
small-drop regime volume and contact angle are both proportional to
the influx.  In the large-drop regime the contact angle approaches a
constant (or increases with growing influx following a power law with
an exponent smaller than $1/5$), whereas the volume depends
quadratically on influx. The latter is easily explained noticing that
the evaporative outflux for large drops is proportional to the surface
``area'' of the drop (negligible influence of Laplace and disjoining
pressure). For a constant contact angle the area under the parabola
depends quadratically on its arc length.  For the influx to balance
the outflux, the surface area has to grow proportionally with the
influx, i.e., the volume increases quadratically with the influx.

Inspecting Fig.~\ref{fig:vol} further, one notices that the overall
behaviour is different for larger ($\epsilon\gtrsim10^{-3}$) and smaller
($\epsilon\lesssim10^{-3}$) drops.  In the former case the transition between
the small-drop and large-drop regime is monotonic, i.e., the slopes of
the curves in Fig.~\ref{fig:vol} change monotonically. In contrast,
for small $\epsilon\lesssim10^{-3}$ in the transition range one may
define a third region where the slopes of the $V(q_0)$ and
$\theta_\mathrm{app}(q_0)$ curves pass through a maximum.

The tendency towards a constant contact angle for increasing volume
can also be observed in Figs.~\ref{fig:vol2}(a) and (b) where we plot
the drop volume and the apparent contact angle, respectively, as a
function of the length scale ratio $\epsilon$ for various fixed
influxes for rather large drops. We find that for large drops,
  the volume as well as the contact angle decrease for increasing
  length scale ratio $\epsilon$ roughly as $\epsilon^{-1/4}$. This
  agrees with the asymptotic expression
  determined above in Section~\ref{sec:asymp}.

For smaller drops, deviations from the power law are found at larger
$\epsilon$. Interestingly, the dependence of the contact angle on
$\epsilon$ seems to approach a limiting curve for large
drops. In the following, we employ the curve for the largest
  drops in Fig.~\ref{fig:prof2} as an approximation to the asymptotic
  dependence of $\theta_\mathrm{app}$ on $\epsilon$ for infinitely
  large drops.

\section{Time evolution without influx}
\mylab{sec:timesim}

Next, we study the time evolution of evaporating droplets without
influx through the substrate, i.e., we simulate Eq.~(\ref{film:dl})
with $q(x)=0$. The domain size $D$ and boundary conditions at $x=0$
and $x=D$ correspond to the ones used in the steady state calculations
in the previous section.  We use three different initial profiles
$h_i(x)=h(x,t=0)$ of equal maximal height $h_m$ and volume $V$: (i) a
parabola $h_i(x)=(h_m-1)(1-x/x_c)^2+1$ with $x_c=3V/(h_m-1)$ for $0\le
x\le x_c$ and $h_i(x)=1$ for $x>x_c$; (ii) a gaussian
$h_i(x)=(h_m-1)\exp((x/\sigma)^2)+1$ with
$\sigma=2V/\sqrt{\pi}(h_m-1)$; and (iii) the steady-state solution of
identical $V$ and $h_m$ obtained in section~\ref{sec:ssd}.

\begin{figure}[h]
\includegraphics[width=0.7\hsize]{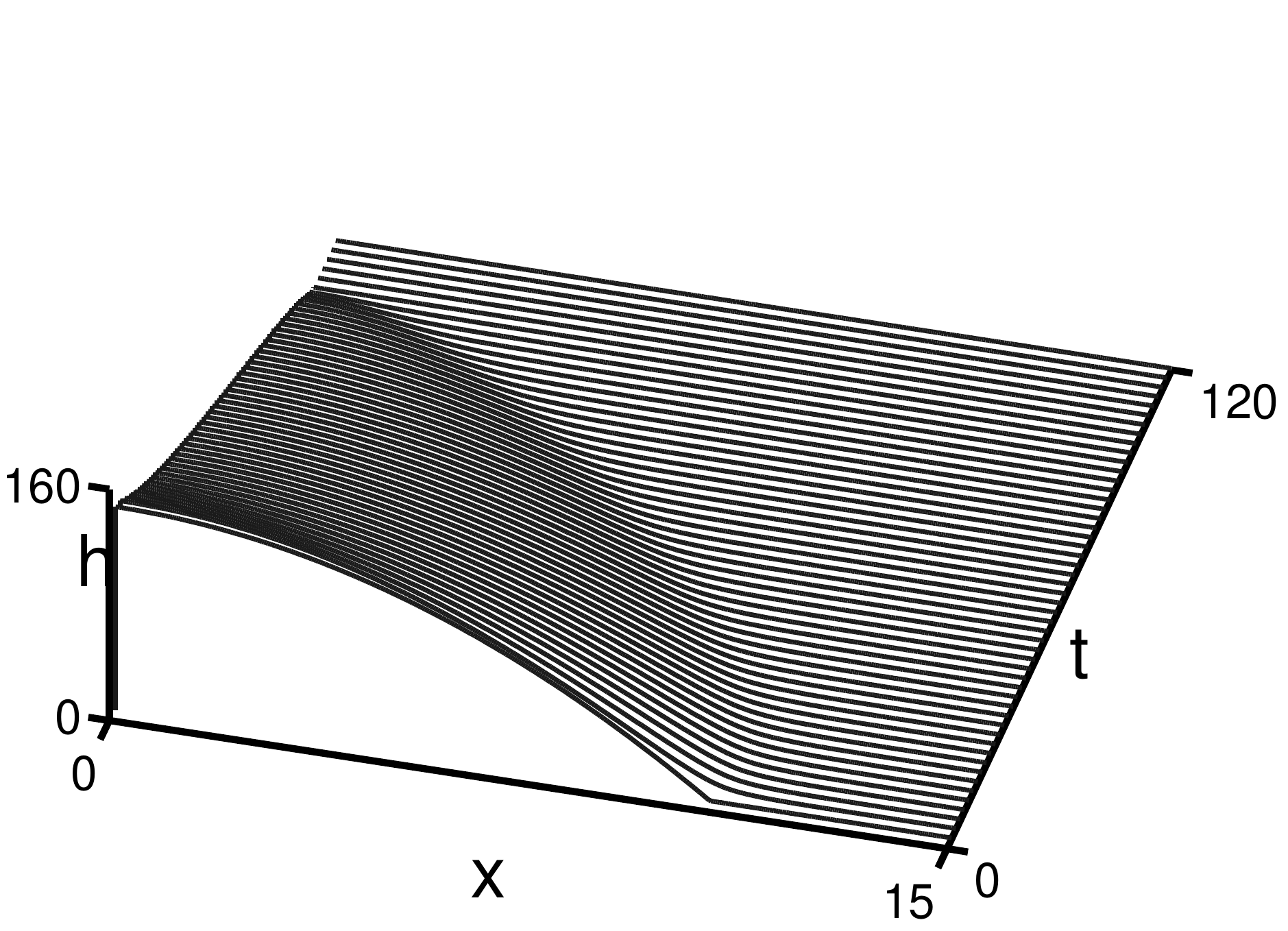}\\
\caption{Space-time plot of an evaporating droplet for
  $\epsilon=10^{-4}$. The initial profile is a
  parabola on a precursor film of thickness $h_p=1$. It has a volume
  of $V=1000$ and maximal height of $H_\mathrm{max}=140.6$. 
  The corresponding contact angle is $\theta_\mathrm{ini}=26.3$.  The
  initial height corresponds to the one at $V=1000$ for the steady
  state drops with influx for the corresponding $\epsilon$ (obtained
  in section~\ref{sec:ssd}).  }
\mylab{fig:st1}
\end{figure}

\begin{figure}[h]
\includegraphics[width=0.5\hsize]{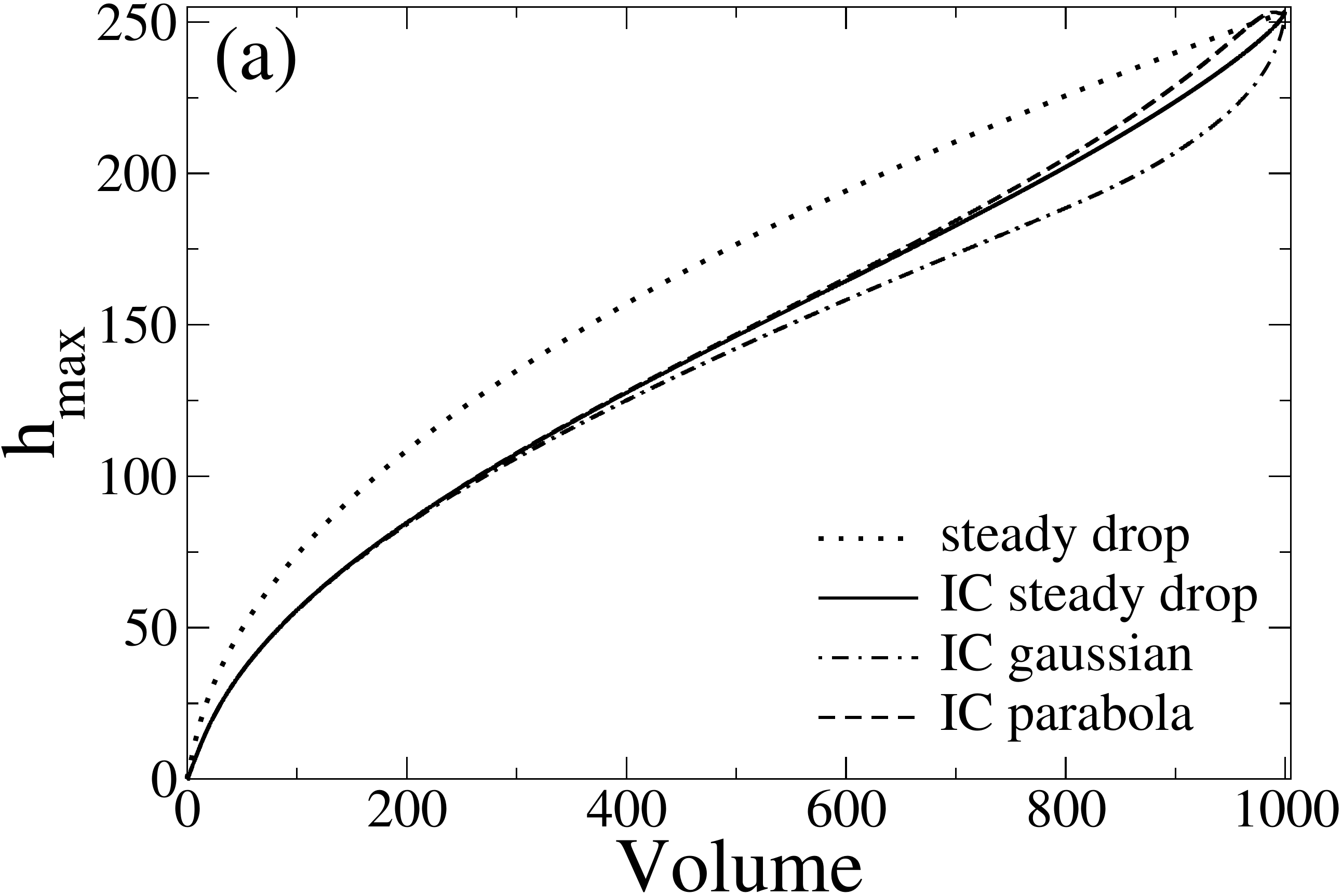}
\includegraphics[width=0.5\hsize]{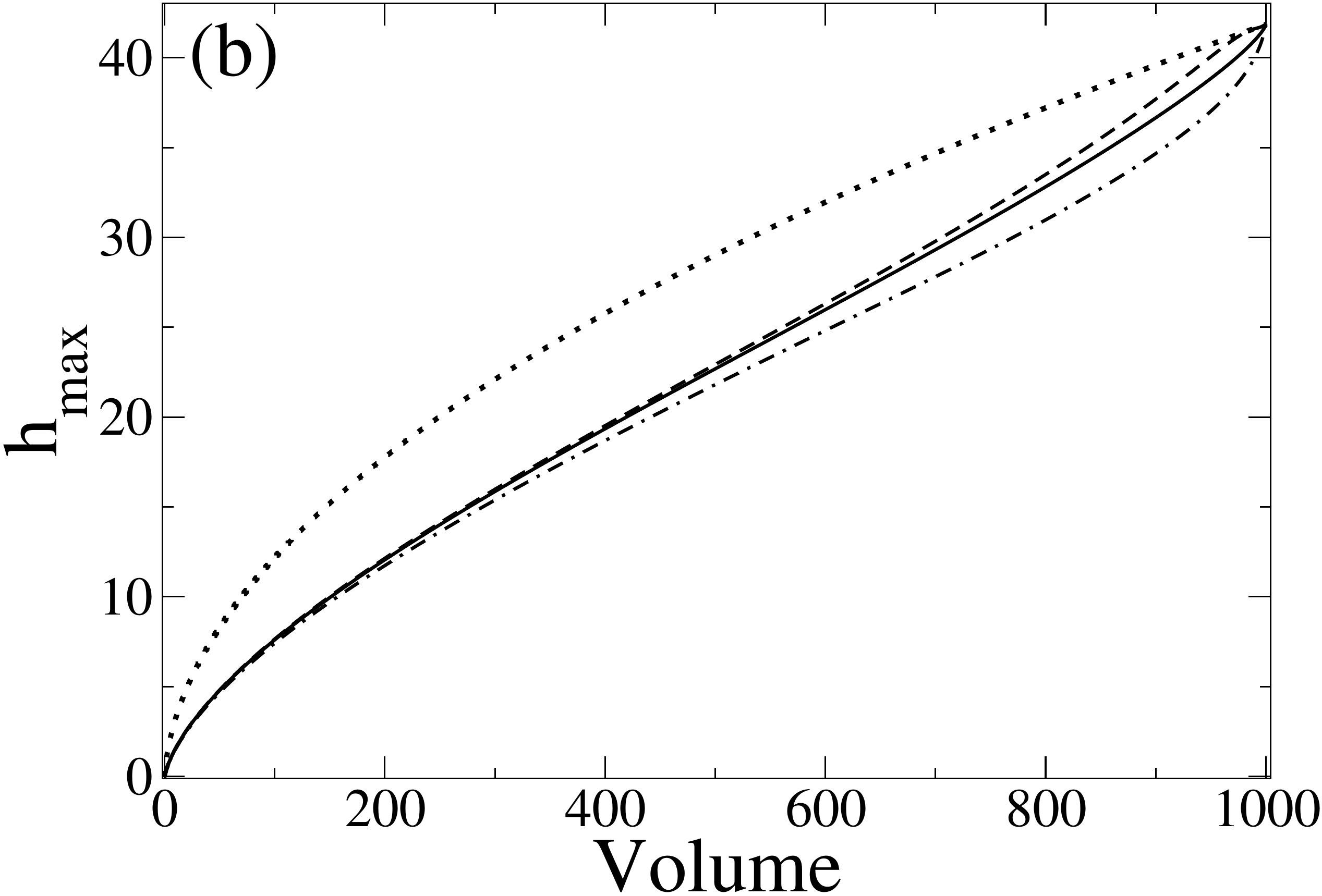}
\caption{Trajectories in the phase plane spanned by the maximal drop
  height and drop volume for (a) $\epsilon=10^{-6}$ and (b)
  $\epsilon=1$.  Shown are curves resulting from (i) time evolutions
  for three different initial profiles of equal maximal height and
  volume (parabola, gaussian and steady state with influx), and (ii)
  calculations of steady state solutions with influx as obtained by
  continuation.}
\mylab{fig:pp-volmax}
\end{figure}

Fig~\ref{fig:st1} shows a space-time plot for a typical time evolution
observed when using an initial parabolic drop profile that has the
same height and volume as a fed drop obtained in
section~\ref{sec:ssd}. The case shown is for $\epsilon=10^{-4}$. At
early times the contact line region relaxes under the influence of the
disjoining pressure, thereby decreasing the apparent contact
angle. Subsequently, the width and height of the drop decrease
monotonically until at about $T=100$ the drop has vanished and only
the stable precursor film remains. When starting (as in the present
case) with the drop measures (volume and height) as obtained for the
drop with influx, the evolution always looks similar. In particular,
we have not found that the drop macroscopically spreads at the
beginning (by ``macroscopic'' we mean a spreading that goes beyond the
small local relaxation at the contact line).

\begin{figure}[h]
\includegraphics[width=0.7\hsize]{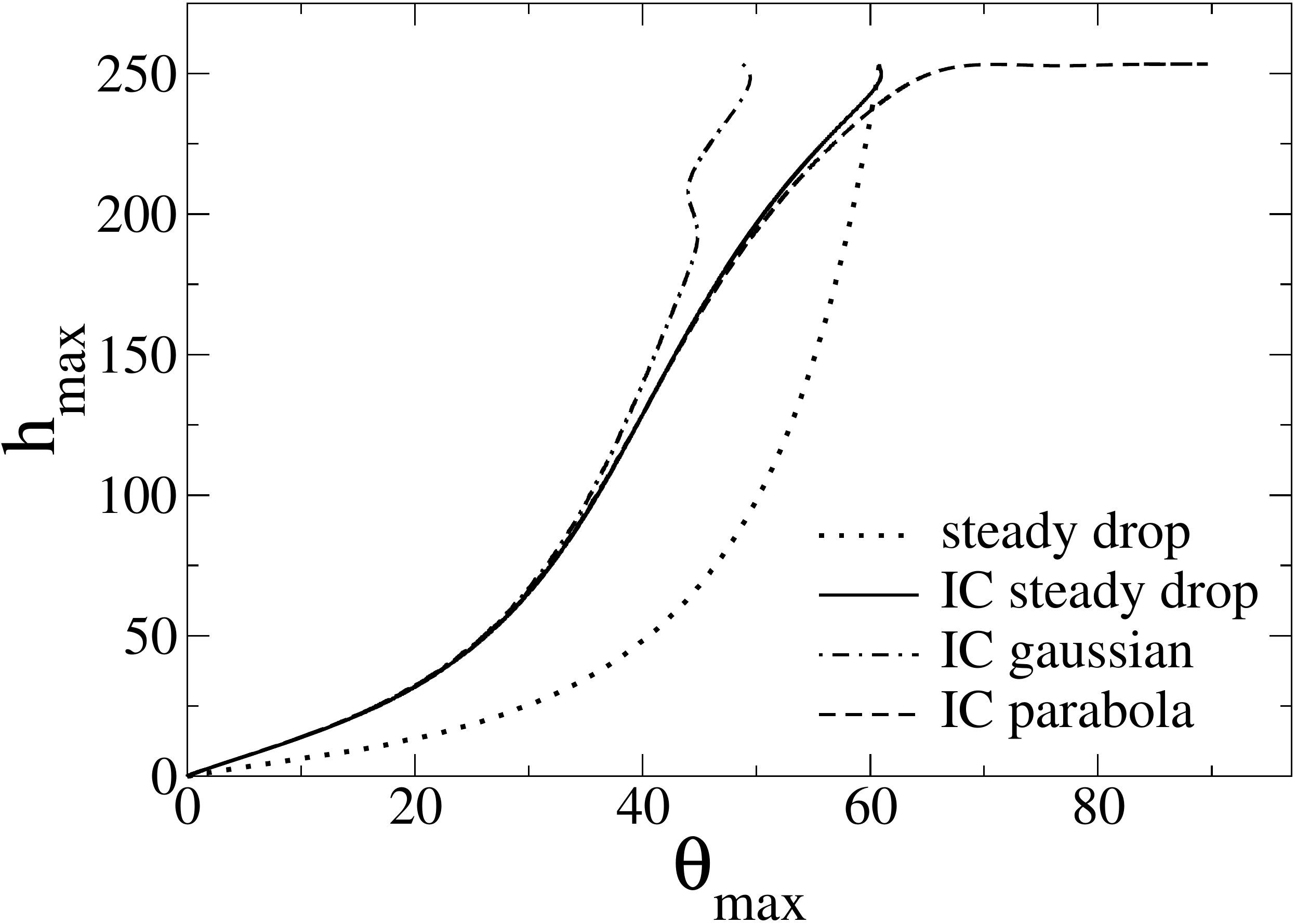}%
\caption{Trajectories in the phase plane spanned by the maximal drop
  height and apparent contact angle for $\epsilon=10^{-6}$.
  Cases shown correspond to the ones in
  Fig.~\ref{fig:pp-volmax}(a).}
\mylab{fig:pp-anglemax}
\end{figure}

\begin{figure}[h]
\includegraphics[width=0.7\hsize]{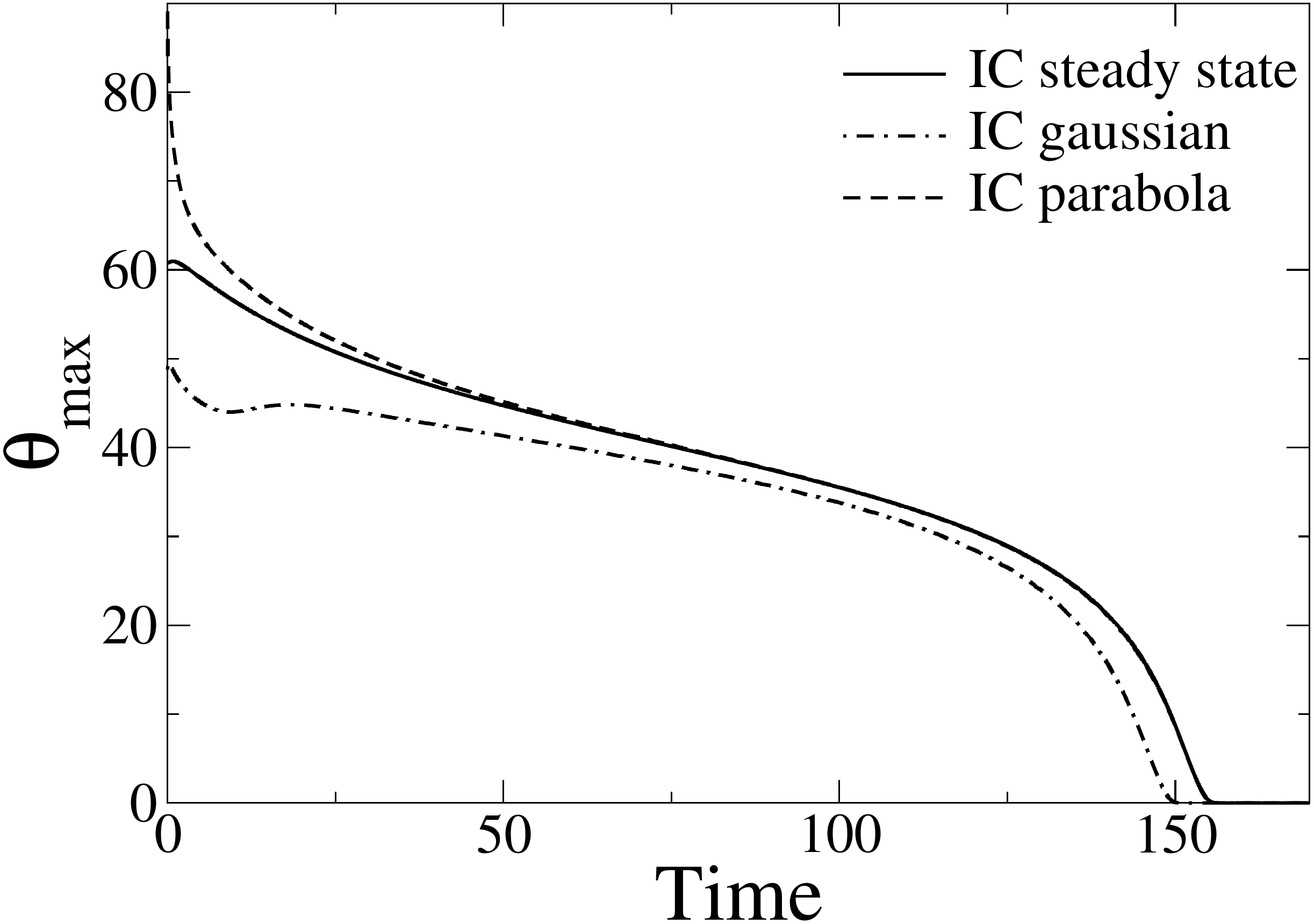} 
\caption{Shown is the dependence of the apparent contact angle on
  time for $\epsilon=10^{-6}$. 
  The given
  cases correspond to the three evaporating drops in
  Fig.~\ref{fig:pp-volmax}(a).}
\mylab{fig:angletime}
\end{figure}

\begin{figure}[h]
\includegraphics[width=0.45\hsize]{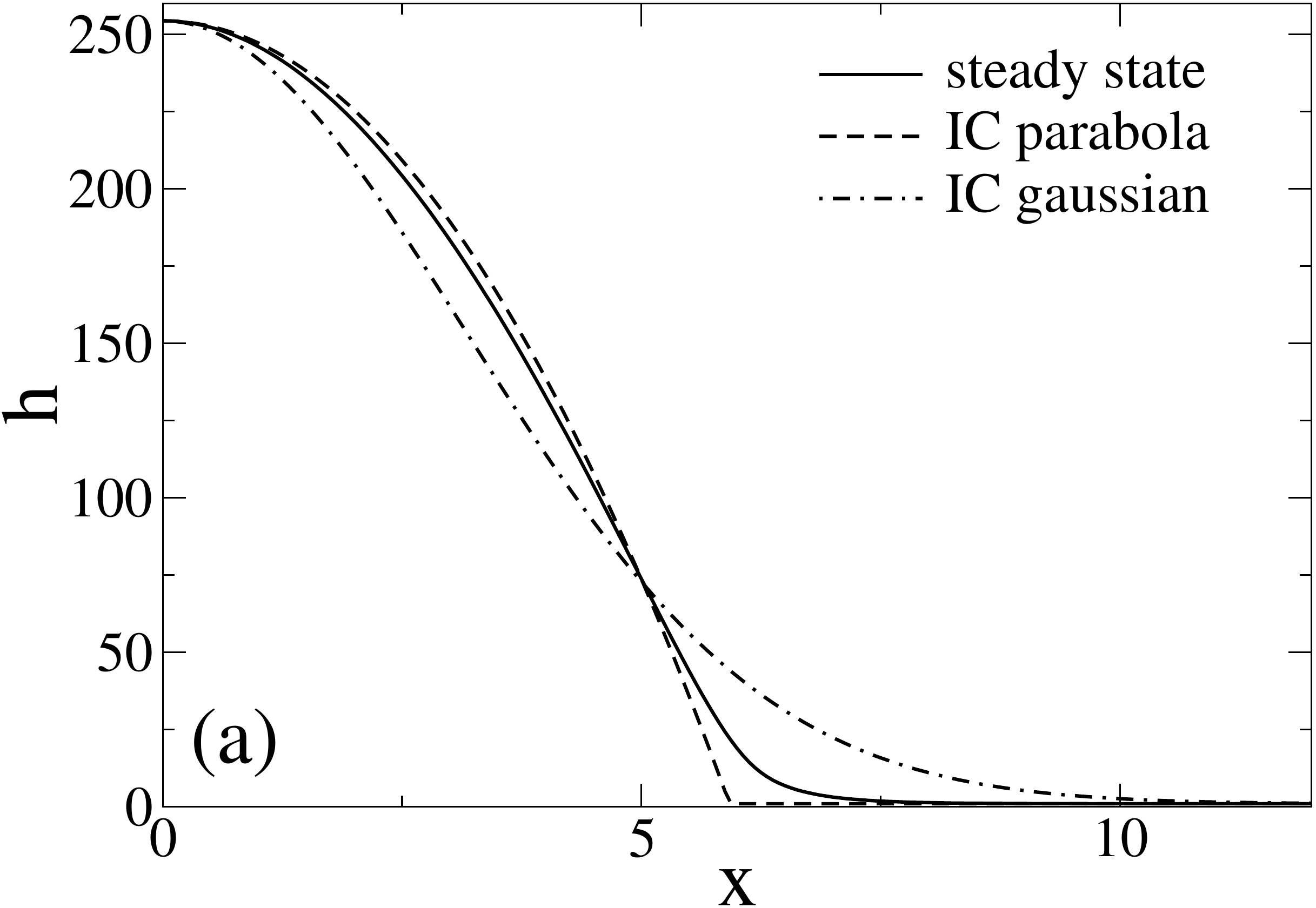}
\includegraphics[width=0.45\hsize]{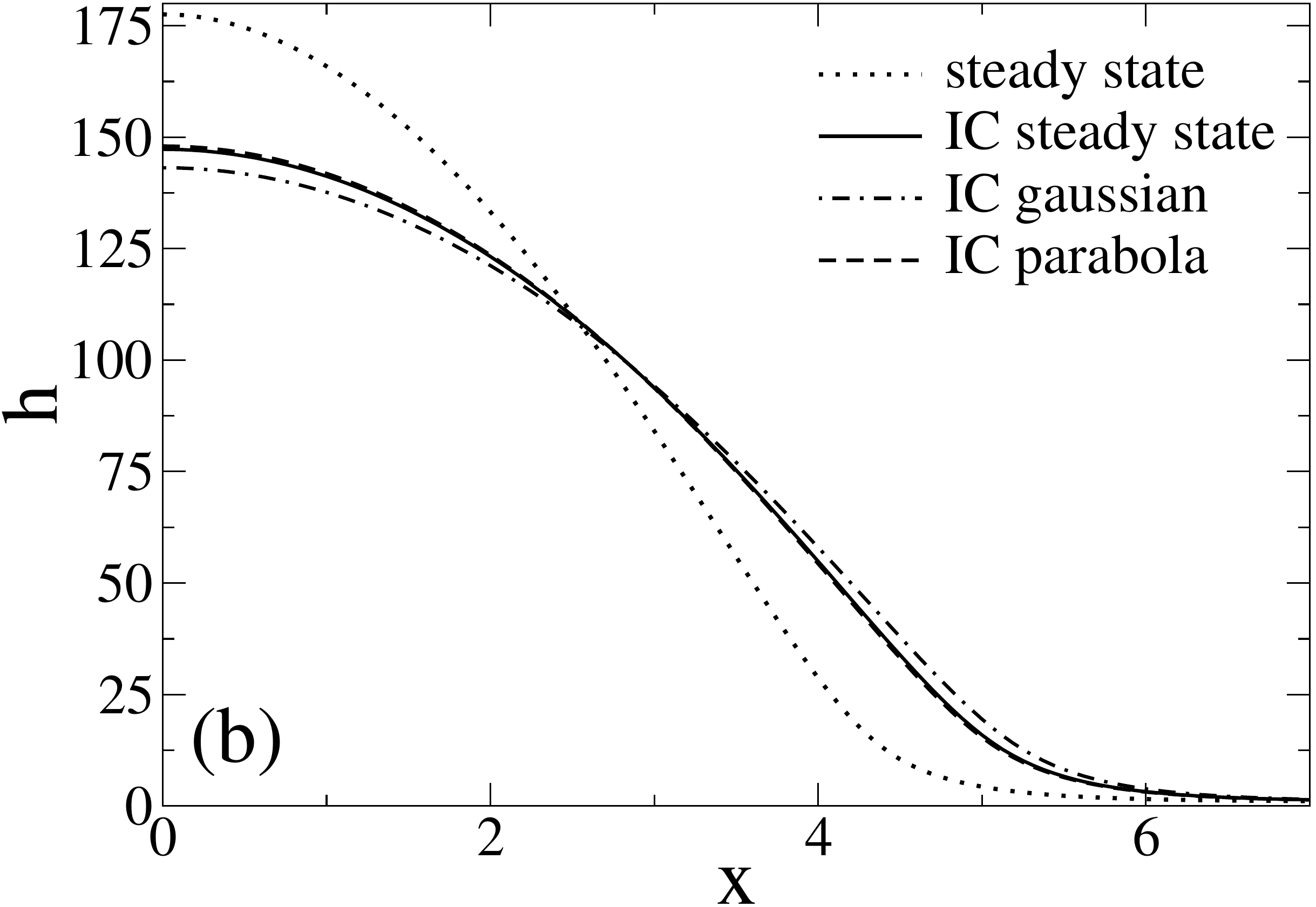} \\ 
\includegraphics[width=0.45\hsize]{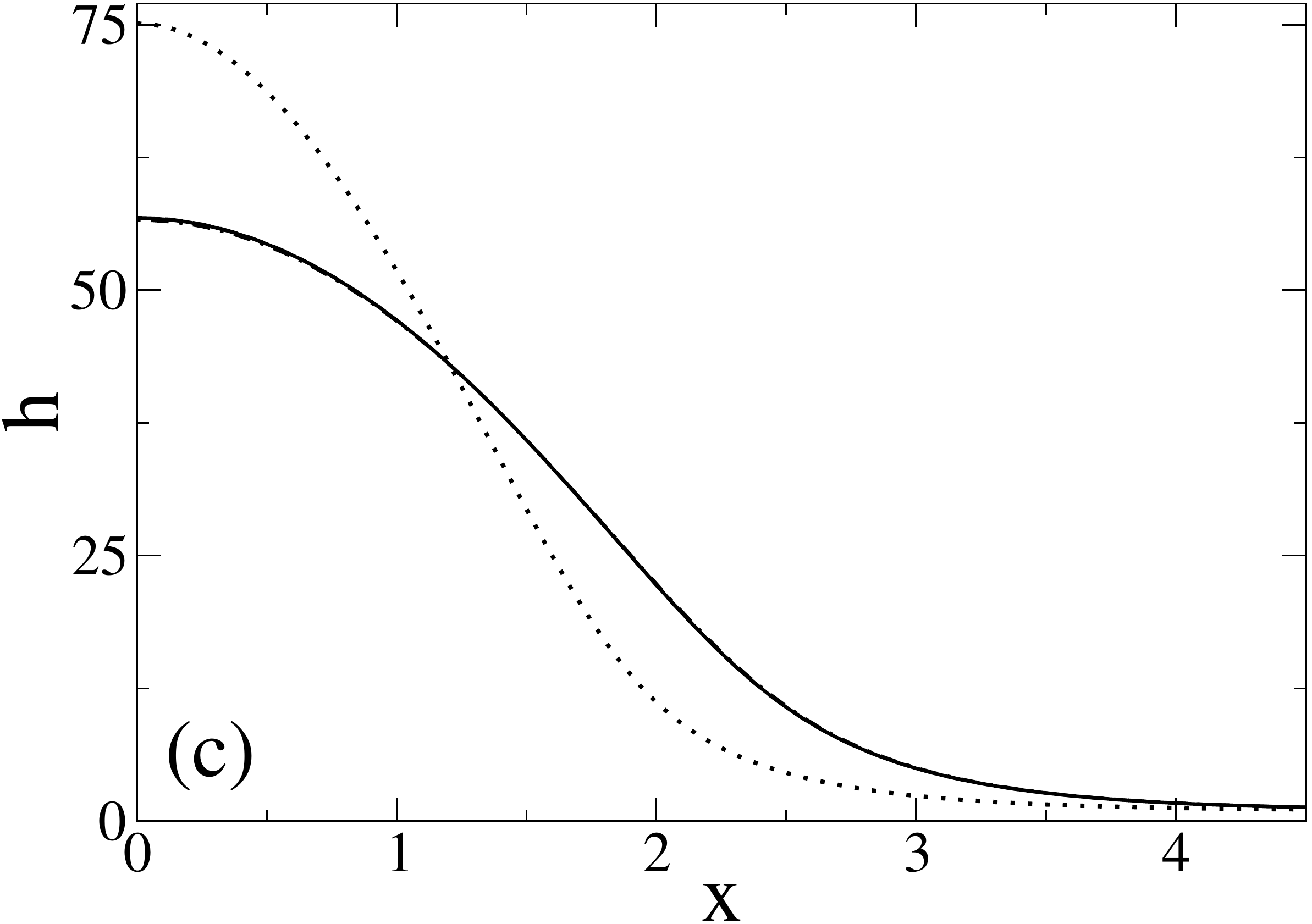}
\includegraphics[width=0.45\hsize]{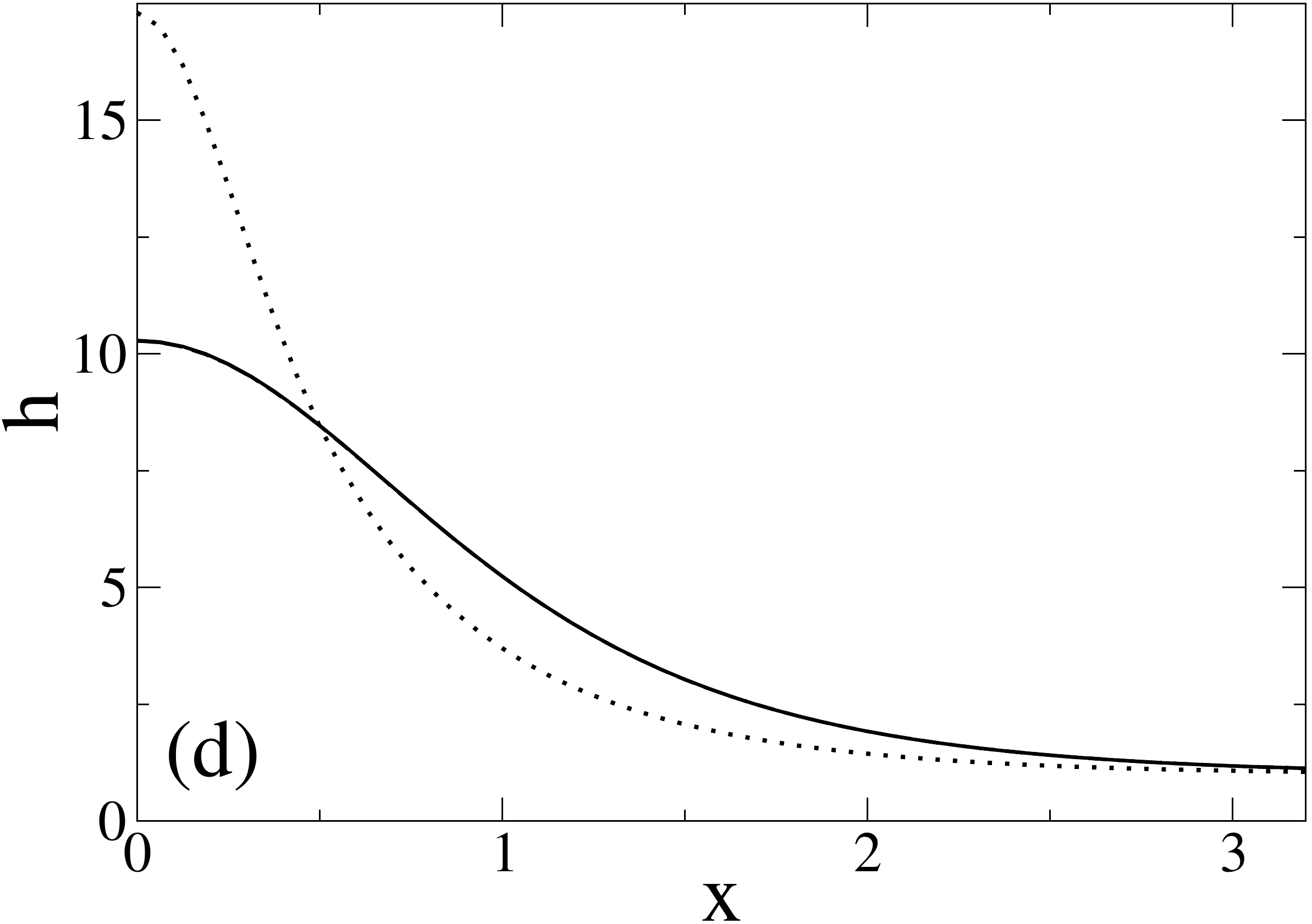}
\caption{Given are for $\epsilon=10^{-6}$ drop profiles for selected
  drop volumes during the course of evaporation (for the three
  different initial profiles). We show as well the steady state drop
  of the same volume. Panel (a) gives the initial profiles at $V=1000$
  whereas panels (b) to (d) show profiles at $V=500$, $V=100$, and
  $V=10$, respectively.  }
\mylab{fig:prof-evol1}
\end{figure}

A more complete picture of the time evolution for different initial
profiles is obtained by considering the dependence of overall measures
on time, and the trajectories of time evolutions in various ``phase
planes''.  For the axes of the latter we choose measures that do not
change when the domain size is varied for an identical drop.  We give
results in two such phase planes, namely, the one spanned by volume
and maximal drop height (Fig.~\ref{fig:pp-volmax}) and the one spanned
by maximal drop height and apparent contact angle
(Fig.~\ref{fig:pp-anglemax}).  The change of the contact angle over
time is given in Fig.~\ref{fig:angletime}, whereas
Fig.~\ref{fig:prof-evol1} shows selected drop profiles.
Figs.~\ref{fig:pp-volmax}(a) and (b) compare results for very small
$\epsilon=10^{-6}$ and the largest used $\epsilon=1$. As the results
are qualitatively rather similar, the remaining figures
~\ref{fig:pp-anglemax} to ~\ref{fig:prof-evol1} are for
$\epsilon=10^{-6}$ only.

Scrutinising Figs.~\ref{fig:pp-volmax} to~\ref{fig:prof-evol1} one
makes several observations: (i) The time evolutions starting from the
three different initial profiles converge after some initial
adjustments whose details depend on the particular initial profile
shape.  (ii) The convergence is slightly faster for smaller
$\epsilon$. Here, ``faster'' means that the trajectories approach each
other at higher volume [cf.~Fig.~\ref{fig:pp-volmax}]. In absolute
terms the overall evolution becomes faster with increasing $\epsilon$.
Thereby, the trajectories of the initial parabola and the initial
profile taken from the steady state calculations approach each other
earlier than they are approached by the trajectory of the initial
gaussian. (iii) The family of steady profiles with influx represents
drops clearly distinct from the ``freely evaporating'' shrinking
drops. The family of steady profiles does not approach the
trajectories of evaporating profiles when the drops become small.
Even for very small drops their contact angle remains always larger by
a roughly constant factor than the one of the evaporating drops. The
factor is about two for $\epsilon=10^{-6}$ and approaches four for
$\epsilon=1$. (iv) The overall picture in Fig.~\ref{fig:pp-volmax} for
different $\epsilon$ looks very similar, only the $h_\mathrm{max}$
axes scale differently. A similar observation holds for the
representations as given in Figs.~\ref{fig:pp-anglemax}
to~\ref{fig:prof-evol1} where, however, both axes would need to be
scaled.

Next, we discuss the behaviour of the different initial drop profiles
at early times. We use Fig.~\ref{fig:pp-anglemax} as an example.  For
larger $\epsilon$ the behaviour is slightly less pronounced, but all
the curves look qualitatively the same (not shown). For instance, one
obtains a plot that is roughly the one for $\epsilon=1$ when scaling
the $h_\mathrm{max}$-axis and $\theta_\mathrm{app}$-axis of
Fig.~\ref{fig:pp-anglemax} by factors of 1/6 and 1/40, respectively.
A similar rule applies to Fig.~\ref{fig:angletime}, when additionally
scaling time by about 1/5.
In Figs.~\ref{fig:pp-anglemax} and~\ref{fig:angletime}, one finds for
the initial parabola profile a strong decrease in the apparent
contact angle at early times. This corresponds to an adjustment
of the contact line region to the influences of the disjoining
pressure.  As the central drop region nearly coincides with the
initial steady profile (per definition at same volume and height) the
two curves approach each other rather fast. In the course of the time
evolution the central part of the profile remains a parabola.
However, for the gaussian at early times the contact angle changes
non-monotonically: The profile adjusts on the one hand its contact
line region to the disjoining pressure influences (related to the
``earlier wiggle'' in the curve for the parabola in
Fig.~\ref{fig:pp-anglemax}). On the other hand, its central region
adapts to a parabola (second ``wiggle'' in the curve in
Fig.~\ref{fig:pp-anglemax}).

All three profiles approach each other after the initial
adjustments. Their central part can be well fitted by a parabola,
e.g., for $V=500$ [$V=100$] and $\epsilon=10^{-6}$ down to thicknesses
of about $h=60$ [$h=35$]. Keeping the drop volume constant, that
thickness decreases with increasing $\epsilon$ and vice versa.
In contrast, the steady profile of the drop with influx can be fitted
by a parabola in a smaller central part of the drop. The deviation
from the parabola becomes clearly visible, e.g., for $V=500$ [$V=100$]
and $\epsilon=10^{-6}$ at about $h=120$ [$h=60$] (already 20-30\%
below the maximum).  This percentage range remains roughly the same
when changing $\epsilon$ for fixed drop volume.

Our direct comparison of evaporating steady state drops with influx
and evaporating drops without influx shows that the former cannot be
used to approximate the latter as their shapes always differ. A freely
evaporating shrinking drop has always a smaller apparent contact
angle than the steady fed drop. This has been shown for a wide range
of length scale ratios from $\epsilon=10^{-6}$ to $\epsilon=1$. Note,
however, that the differences slowly decrease for decreasing
$\epsilon$.

\begin{figure}[h]
(a)\includegraphics[width=0.45\hsize]{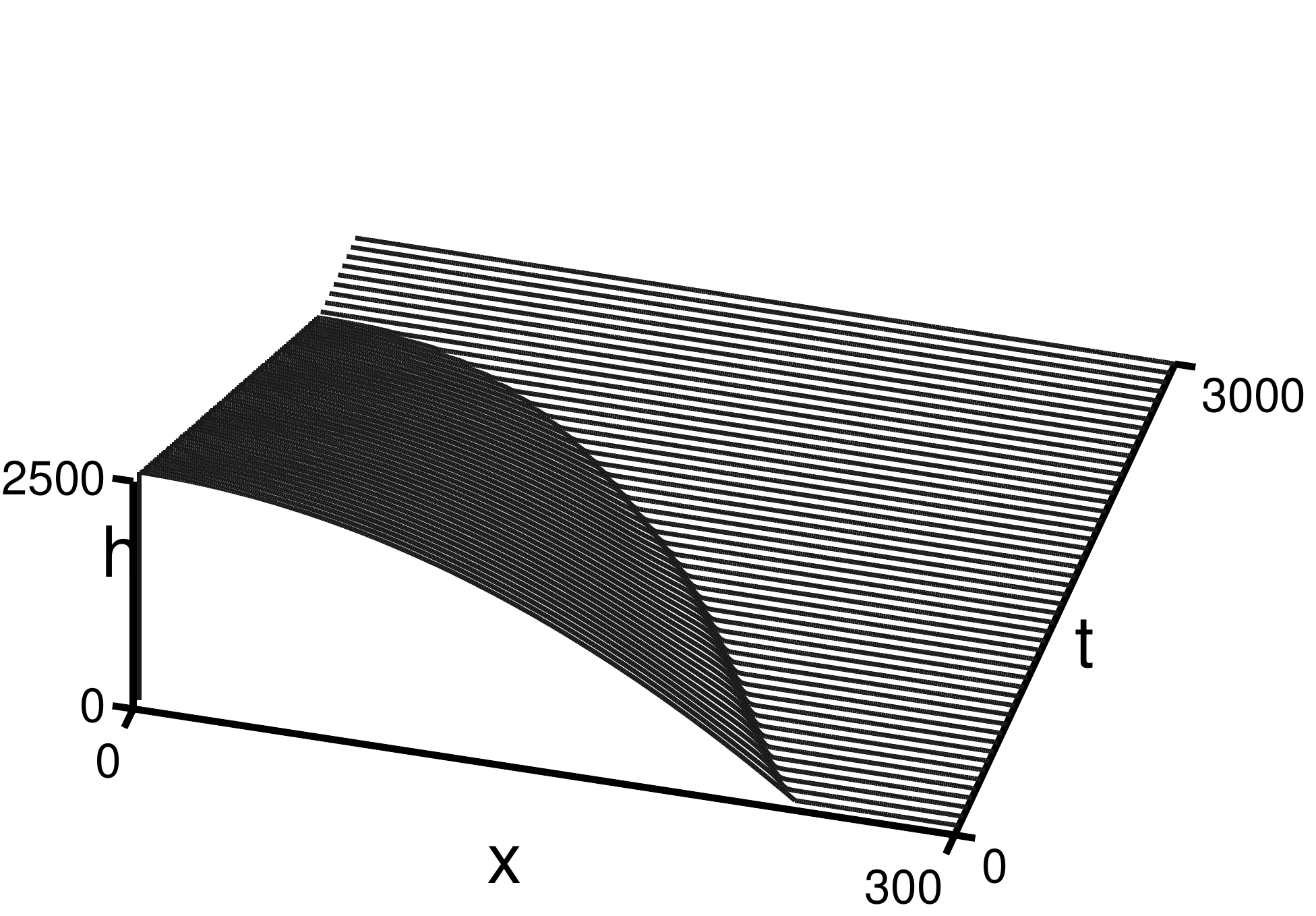}
\includegraphics[width=0.45\hsize]{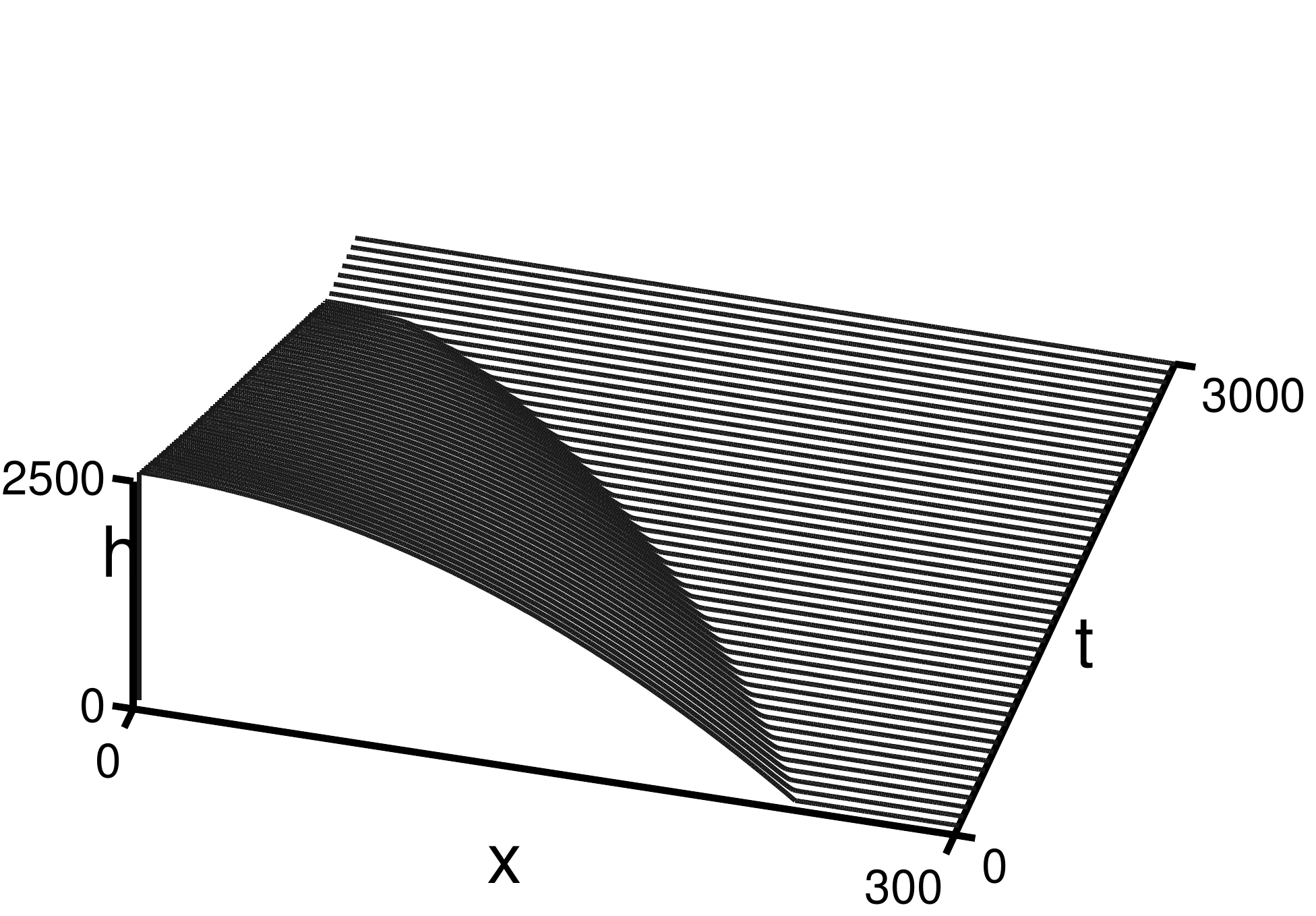}(b)\\
(c)\includegraphics[width=0.45\hsize]{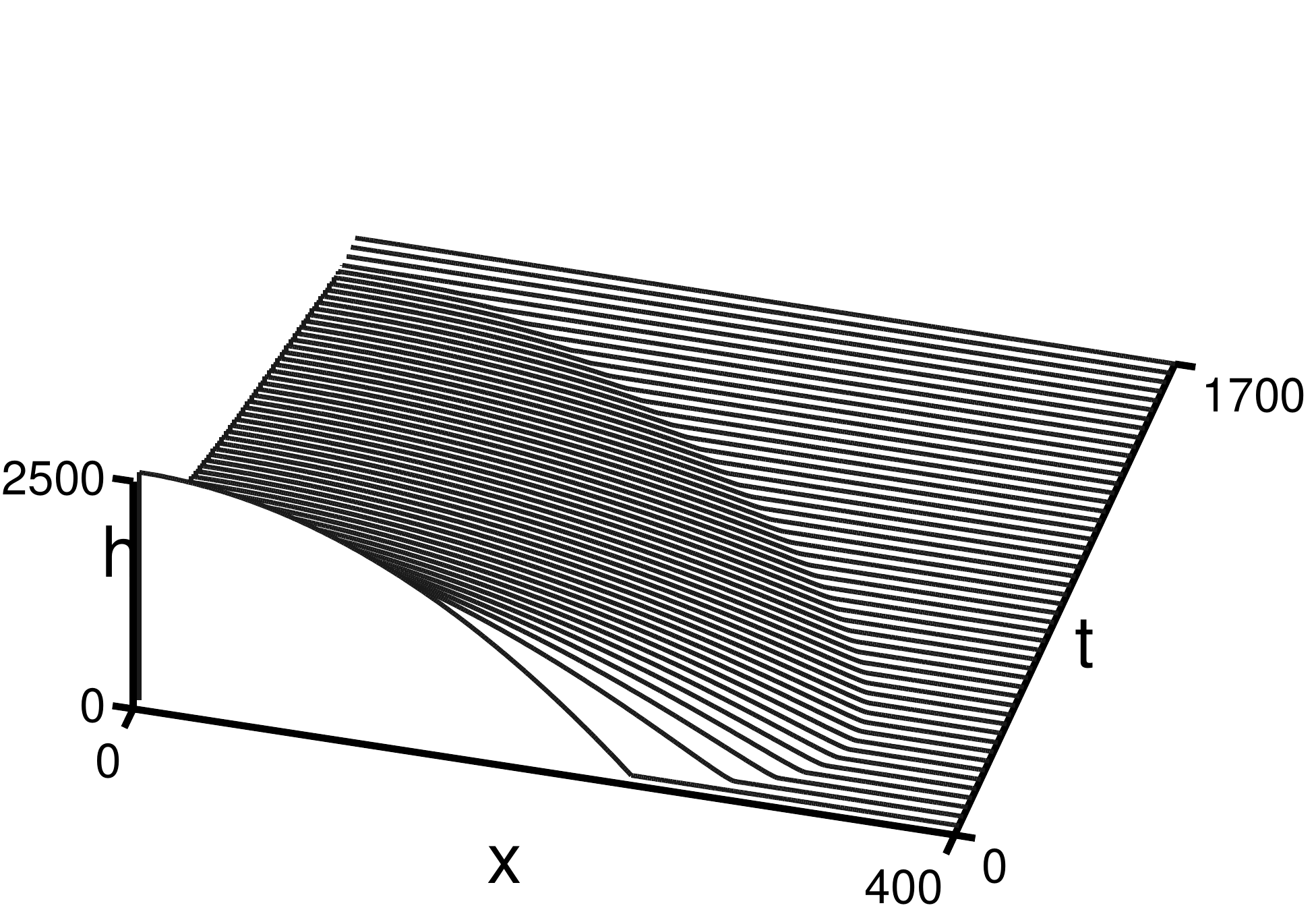}
\includegraphics[width=0.45\hsize]{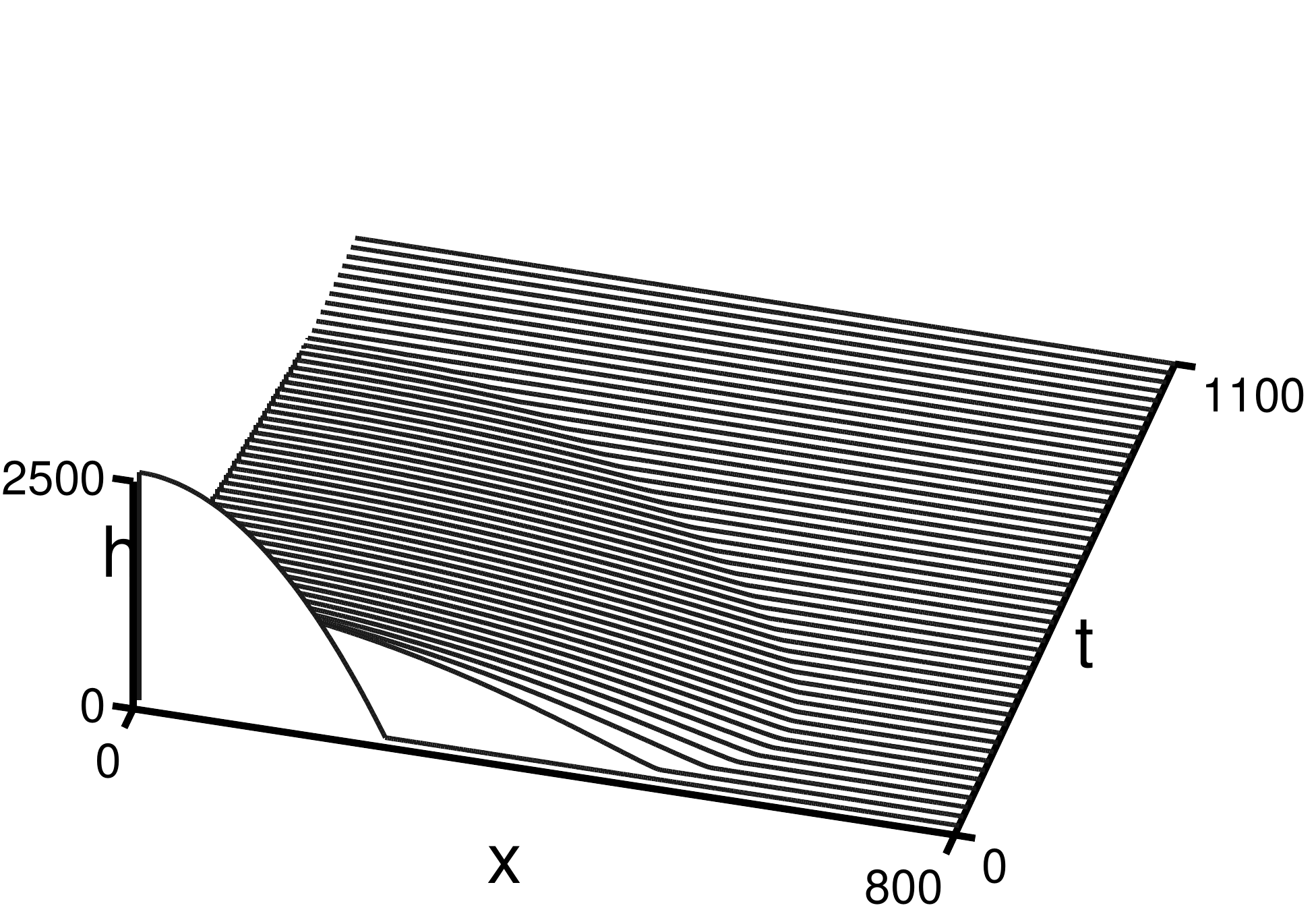}(d)
\caption{Space-time plots of an evaporating droplets for (a)
  $\epsilon=10^{-6}$, (b) $\epsilon=10^{-4}$, (c) $\epsilon=10^{-2}$,
  and $\epsilon=1.0$.  The initial profile is always the same parabola
  of maximal height $H_\mathrm{max}=2500$ and volume $4\times10^{5}$ on a
  precursor film of $h=1$. The corresponding contact angle is
  $\theta_\mathrm{ini}=20.8$.}
\mylab{fig:prof-large}
\end{figure}

We have observed that freely evaporating drops with a similar initial
geometry (volume and height) as per steady-state drops, for the
range of $\epsilon$ explored, never spread macroscopically before their
contact line recedes. However, an initial spreading phase is often
observed in experiments \cite{CBC02,SRAB06,Bonn09}.
To investigate
this further we perform a number of simulations starting with large
parabolic drops. Fig.~\ref{fig:prof-large} gives a set of space time
plots obtained for different length scale ratios from
$\epsilon=10^{-6}$ to $\epsilon=1$. All of them start from the
identical initial profile.  For small $\epsilon\lesssim10^{-3}$
[panels (a) and (b)] the behaviour is very similar to the one
described above for drops with the same initial geometry as the
steady drops: the drops shrink monotonically, their height and width
decrease slowly. However, at larger $\epsilon\gtrsim10^{-3}$ [panels
(c) and (d)] the behaviour is qualitatively different: At early times
the drops spread. Thereby they loose height and gain width quite fast,
the apparent contact angle decreases strongly. Then the drop reaches
a maximal width before the contact line starts to recede again. In the
shrinking stage, the height and width of the drop decrease slowly as
before. The spreading is faster for larger $\epsilon$
[Fig.~\ref{fig:prof-large}(d)].

The contact angle for the initial profile is in all cases
$\theta_\mathrm{ini}=20.8$.  Comparing this with Fig.~\ref{fig:vol2}(b) one
notices that this angle roughly coincides with the limiting contact
angle (for large drops) for $\epsilon\approx3\times10^{-4}$. This
value lies between the regions (in $\epsilon$) where we find receding
and spreading evaporating drops, respectively. Extrapolating from this
finding, we formulate the hypothesis that the steady state drops with
influx studied above in section~\ref{sec:ssd} represent limiting
solutions between the case of spreading and shrinking freely
evaporating drops (without influx).

\begin{figure}[h]
\includegraphics[width=0.7\hsize]{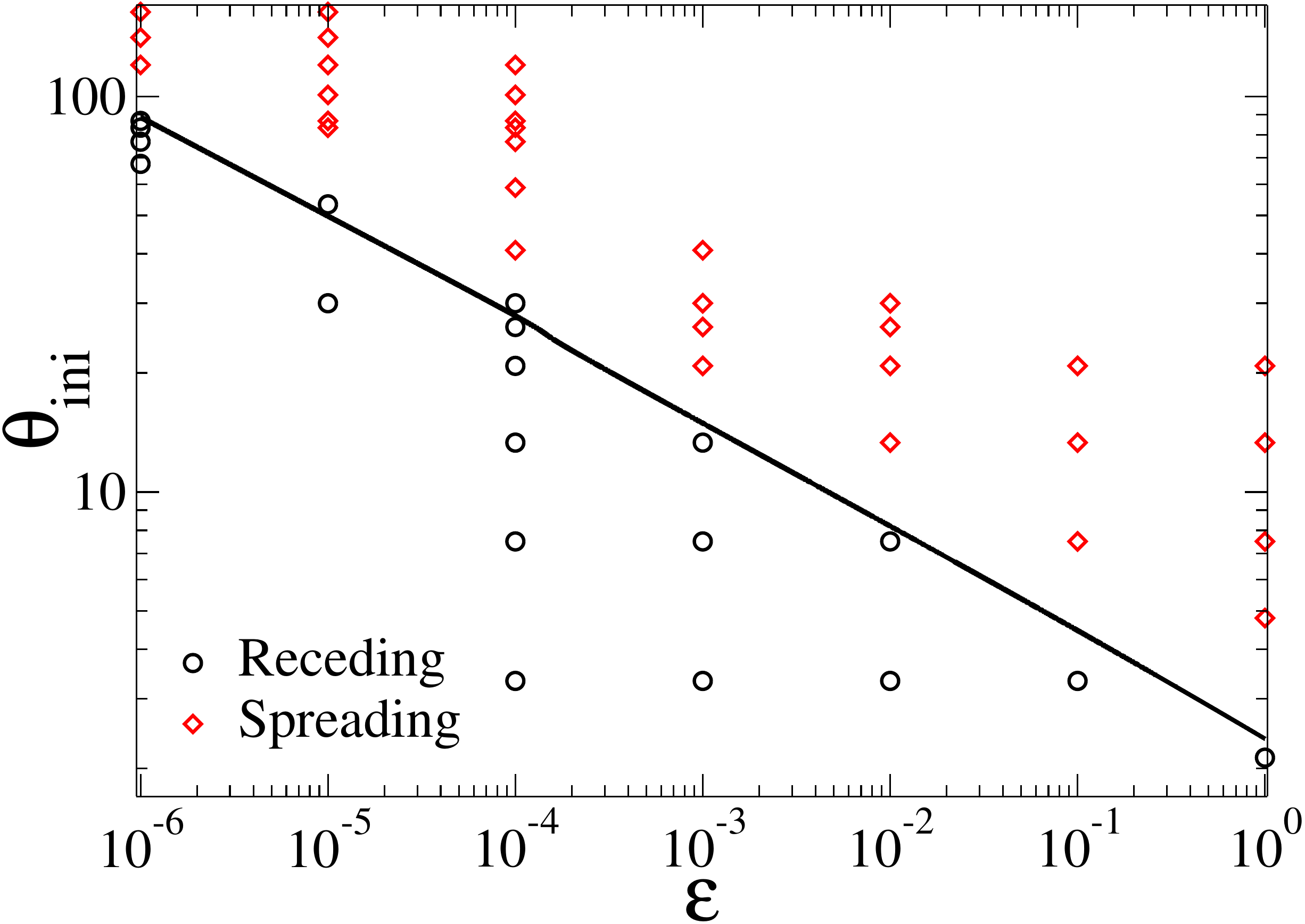}
\caption{(colour online) Phase diagram indicating the initial behaviour
  of an evaporating drop. In the plane spanned by the initial contact
  angle $\theta_\mathrm{ini}$ and the length scale ratio $\epsilon$ we
  indicate where the drop initially spreads, and where the contact
  line recedes right from the beginning. Each symbol corresponds to a
  time simulation. The solid line corresponds to the numeric result
  characterising the large steady drops with influx
  (cf.~Fig.~\ref{fig:vol2}(b)).}
\mylab{fig:evap-phase}
\end{figure}

To test the hypothesis we perform a number of time simulations with
parabolic initial drops of different initial contact angle and at
different $\epsilon$. All of them are of the same (large) volume. The
results are given as a scatter plot in Fig.~\ref{fig:evap-phase}
together with the curve for the large steady drops obtained in
section~\ref{sec:ssd} [solid line of Fig.~\ref{fig:vol2}(b)]. For each
initial condition we record whether the drop spreads initially or
directly starts to recede.
Our results indicate that the above hypothesis seems to hold.  The
transition between initial spreading for large initial contact angle
and a receding of the contact region right from the beginning roughly
coincides with the power law dependence $\theta\sim\epsilon^{-1/4}$
(Fig.~\ref{fig:vol2}(b), curve for large volume) not only in the power
but as well in the prefactor. The prefactor of the curve obtained from
the steady drops with influx seems to be slightly below the transition
found in the time simulations. Further studies will be necessary to
give a more detailed account.

\section{Conclusions}
\mylab{sec:disconc}

We have analysed a thin film evolution equation for a wetting
evaporating liquid on a smooth solid substrate.  We have focused on
slowly evaporating small sessile droplets where thermal effects are
insignificant. Employing the model, we have first studied evaporating
drops as steady state solutions for the case when they are fed through
a porous part of the substrate. In particular, an asymptotic analysis
has focused on the transition region between the precursor film and
the bulk drop; and a numerical continuation of steady state
  drops has determined the fully non-linear drop profiles as a
  function of the overall influx for various values of the length
  scale ratio $\epsilon$. We have found that for large steady drops,
the volume as well as the apparent contact angle decrease for
increasing $\epsilon$ roughly as $\epsilon^{-1/4}$. This does
  well agree with the scaling
  $\epsilon^{-1/4}$
  determined via the asymptotic analysis. Note that the mentioned
  logarithmic corrections to both the overlap range where the matching
  is done and the resulting profile slope (apparent contact angle) are
  found to have opposite effects and are too subtle for
  order-of-magnitude estimates.

Furthermore, we have employed the model to study the time
evolution of freely evaporating drops that are not fed through the
substrate, i.e., the full evolution equation has been numerically
integrated. Thereby the time evolution of several different initial
drop shapes (for identical maximal height and volume) has been
compared. It has been shown that freely evaporating drops with
different initial profiles converge onto certain trajectories in phase
space. However, a direct comparison of the freely evaporating drops
without influx with the evaporating steady state drops with influx has
shown that the latter cannot be used to approximate the former. A
freely evaporating shrinking drop has always a smaller apparent
contact angle than the steady drop with influx. Here it has been
investigated for a wide range of length scale ratios from
$\epsilon=10^{-6}$ to $\epsilon=1$. However, as the differences
between the two types of profiles slowly decrease with decreasing
$\epsilon$, further studies should scrutinise the case of even smaller
$\epsilon$.

We have noted that in our simulations the freely evaporating drops
with a similar initial geometry (volume and height) to
steady-state drops with influx, never spread macroscopically before
their contact line starts to recede and the drop shrinks. As drops
undergo an initial spreading phase in many experiments, we have
investigated this further and found that drops spread
[shrink] from the beginning if their initial contact angle is larger
[smaller] than the apparent contact angle of large evaporating drops
with influx.

This seems to be a very promising result that should be further
scrutinised as it might have interesting consequences: (i) If the
apparent contact angle of a steady drop with influx takes the role of
an equilibrium contact angle $\theta_e$, relations between the dynamic
angles and the contact line velocity known from non-volatile partially
wetting liquids \cite{deGe85} could hold. Although, this has recently been
  shown for the case of evaporating partially wetting liquid
  \cite{AGS10}, it remains an open question for the case of a wetting
  liquid that we study
  here.  (ii) It might further be possible to predict the maximal
drop radius and the contact angle at which the initial spreading
ceases, and the 'turn around' to the receding motion occurs. It seems
plausible that the profile at turnaround might actually be identical
to the steady drop profile with influx at the same volume. Note that
the freely evaporating drop spreads \textit{and} evaporates, i.e., the
volume at turn-around does not correspond to the initial one.

Note finally, that here we have studied a particular evaporation model
valid for small drops in situations where thermal aspects and the
dynamics in the surrounding gas phase can be neglected.  However, the
non-isothermal models can all be studied with a similar methodology,
i.e., the properties of steady drops with local influx can be determined
and may be employed to gain a deeper understanding of the coupled
transport and phase change processes.

\section{Acknowledgments}
This work has been supported by the European Union under
Grant No. PITN-GA-2008-214919 (MULTIFLOW).

\bibliographystyle{klunum}

\end{document}